\documentclass[aps,prx,twocolumn,groupedaddress]{revtex4-2}

\usepackage{booktabs}
\usepackage{amsmath}
\usepackage{graphicx}
\usepackage{verbatim}
\usepackage{subfigure}
\usepackage[english]{babel}
\usepackage{verbatim}
\usepackage{multirow}
\usepackage{xcolor}
\usepackage[utf8]{inputenc}

\begin{document}

\title{The {R}otational and {S}pin-{R}otational {L}evel {S}tructure of para-H$_{2}^+$ from {H}igh-{R}esolution {M}QDT-{A}ssisted {R}ydberg-{S}tate Spectroscopy}

\author{I. Doran}
\affiliation{Department of Chemistry and Applied Biosciences, ETH Z\"urich, Z\"urich, Switzerland}
\author{M. Beyer}
\affiliation{%
	Department of Physics and Astronomy, LaserLaB, Vrije Universiteit Amsterdam, de Boelelaan 1081, 1081 HV Amsterdam, The Netherlands%
}
\author{F. Merkt} 
\email{frederic.merkt@phys.chem.ethz.ch (F. Merkt)\hfill} 
\affiliation{Department of Chemistry and Applied Biosciences, ETH Z\"urich, Z\"urich, Switzerland}
\affiliation{Department of Physics, ETH Z\"urich, Z\"urich, Switzerland}
\affiliation{Quantum Center, ETH Z\"urich, Z\"urich, Switzerland}

\date{\today}
\begin{abstract}
The structure of the low-lying rotational levels of the X$^+$ $ ^2 \Sigma_g ^+$ ($v^+=0$) vibronic ground state of para-H$_2^+$ has been determined by combining frequency-comb calibrated continuous-wave high-resolution laser spectroscopy of $n$f Rydberg series in the range of principal quantum number $n$ between 28 and 115 and Rydberg-series extrapolation using multichannel-quantum-defect theory (MQDT). The use of accurate quantum-defect parameters obtained from new ab initio calculations enabled the experimental determination of the pure rotational term values of the $N^+= 2$, 4 and 6 rotational levels of H$_2^+$ with sub-MHz accuracy (174.236\,744\,6(77), 575.455\,632\,5(86) and 1191.385\,571(240) cm$^{-1}$, respectively), and of the corresponding spin-rotational coupling constants with an accuracy of better than 100 kHz (42.21(4), 41.26(8) and 40.04(8) MHz, respectively). These values are in agreement with the results of first-principles calculations that include high-order relativistic and quantum-electrodynamics corrections to the level energies. To reach the reported accuracy in the Rydberg series extrapolation, it was necessary to correct for artificial level shifts arising in the MQDT calculations in the vicinity of local perturbations of high-$n$ Rydberg states with a $v^+=0$ H$_2^+$ ion core caused by low-$n$ core-excited Rydberg states, and resulting from approximations in the treatment of the Rydberg-electron energy in the interacting channels.

\end{abstract}
\maketitle

\section{Introduction}
Molecular hydrogen, in its neutral (H$_2$) and ionized (H$_2^+$) forms and their deuterated isotopomers HD$^{(+)}$ and D$_2^{(+)}$, is a fundamental molecular system, with which physical theories and models can be tested by comparison of experiment and theory. First-principles quantum-mechanical calculations treating relativistic and quantum-electrodynamics (QED) effects perturbatively to high orders in the fine-structure constant $\alpha$ provide extremely precise predictions of the level structures, at the level of about 300 kHz and about 1 kHz for the rovibrational levels of the electronic ground state of H$_2$ \cite{puchalski19a,puchalski19b} and H$_2^+$ \cite{korobov06a,korobov08a,korobov06c,korobov14a,korobov17a,koelemeij22a,karr23a,haidar22a}, respectively. Calculations of the hyperfine structure of these ions have also been performed with high accuracy \cite{korobov06c,koelemeij22a,haidar22a}. In recent years, the validity of these calculations has been confirmed experimentally with increasing precision, reaching the level where comparison between theory and experiment helps improving the values of the fundamental constants (e.g., $\alpha$ and the Rydberg constant $R_\infty$) and particle properties (e.g., magnetic moments, nuclear radii, and nuclear-to-electron mass ratios) used as parameters, and even searching for physics beyond the standard model of particle physics \cite{patra20a,alighanbari20a,kortunov21a,kortunov24a,germann21a,delaunay23a,karr23a,schenkel24a, schiller24a}.

H$_2$ and H$_2^+$ also play an important role in studies of primary photo-chemical and -physical processes such as photoionization and photodissociation, but the excited molecular states that are typically involved in these processes cannot currently be calculated as precisely as the electronic ground state. The level structures of H$_2$ and H$_2^+$ are linked through the process of photoionization and this link is elegantly exploited by multichannel quantum-defect theory (MQDT) \cite{seaton83a,fano70a,herzberg72a,greene85a,jungen96a,jungen11a} to relate the neutral and ionized forms of a molecule through the concept of ionization channels. An ionization channel includes the ionization continuum (e.g., H$_2^+(\alpha^+) + {\rm e}^-(\epsilon,\ell)$ in the case of H$_2$) associated with a given quantum state $\alpha^+$ of the ionized species and with a photoelectron partial wave of orbital angular momentum quantum number $\ell$ at energy $\epsilon$ in the continuum, and the infinite series $n\ell[\alpha^+]$ of Rydberg states with principal quantum number $n$ converging on the H$_2^+(\alpha^+) + {\rm e}^-(\epsilon=0)$ threshold \cite{greene85a,jungen11a,merkt11a}. As $n\rightarrow \infty$, the discrete level structure of the neutral molecule approaches that of the ionized molecule and there is therefore no fundamental reason why calculations of electronically highly excited states of neutral molecules could not reach the accuracy achievable for the ionized species. Standard ab initio quantum-chemical methods fail in this respect because of the rapidly increasing size of the basis sets required to reach convergence. In contrast, MQDT, which treats the ionization channels and their interactions as electron-ion collisions in the framework of collision theory, becomes increasingly accurate as $n$ increases because the description of the channels as consisting of an electron and an ion core becomes more and more realistic. 

This convergence enables the determination of the level structure of the ionized species through measurements of Rydberg spectra at high $n$ values. The first experimental determination of the rovibrational level structure of H$_2^+$ in its X$^+$ $^2\Sigma_g^+$ electronic ground state was indeed through measurements of high Rydberg states of H$_2$ and extrapolation using early forms of MQDT \cite{herzberg72a}. Over the years, MQDT has been systematically refined and extended (see, e.g., Refs. \cite{atabek74a,jungen77a,greene85a,du86a,jungen90a,jungen90b,huber90a,huber94a,jungen97a,jungen97b,jungen98a,osterwalder04a,jungen11a,kay11a,sprecher14a,sprecher14x,sommavilla16a}) and can be used today to determine the level structure of ionized molecules with an accuracy limited by the bandwidth of the radiation used to record Rydberg spectra, or the natural linewidth or Doppler broadening \cite{osterwalder04a,cruse08a,beyer19a,doran24a}. Even more importantly, MQDT can also be used to study molecular photoionization and characterize the structure and dynamics of highly excited electronic states of neutral molecules.

In the present article, we report on a systematic investigation, by high-resolution laser spectroscopy and over a broad range of $n$ values, of the $n$f Rydberg series converging to the lowest rotational states ($N^+$ up to 6) of the ground vibronic state of para-H$_2^+$. The purpose of this investigation was threefold.
Firstly, it aimed at testing a new set of purely ab initio quantum defect parameters extracted from recent high-level electronic-structure calculations of low-$n$ Rydberg states \cite{silkowski21a,silkowski22a,silkowski23a,silkowski24a} by comparing the results of full MQDT calculations with those obtained by high-resolution spectroscopy. In the course of this study, we observed, with full spectroscopic detail, a limitation of the currently implemented fully ab initio MQDT formalism arising from the ambiguity in the definition of the Rydberg-electron energy in spectral regions where Rydberg series are strongly perturbed by channel interactions \cite{greene85b,gao90a,hvizdos20a,jungen11a}. We demonstrate how the systematic errors originating from this ambiguity can be strongly suppressed using an empirical two-channel correction model.

Secondly, it aimed at providing a full understanding of the phenomena accompanying the uncoupling of the Rydberg electron from the ion arising at increasing $n$ value. The level structure evolves from a situation, at low-$n$ values, where the Rydberg and core electrons are coupled through the electrostatic exchange interaction and have a well-defined total electron spin to a situation, at high $n$ values, where the core electron is coupled to the ion-core rotation and the Rydberg states have fully mixed singlet and triplet character. Series converging to even $N^+$ rotational levels were selected to specifically investigate the effects of the spin-rotation fine structure of the ion core, which are masked in ortho-H$_2^+$ by the much stronger uncoupling effects caused by the hyperfine interaction \cite{osterwalder04a}. A first analysis of this uncoupling phenomenon by MQDT was reported in an investigation of the $n$f$N^+=2$ series of para-H$_2$ by THz spectroscopy from $n$d levels \cite{haase15a}. This previous study led to the determination of the interval between the lowest two rotational levels of para-H$_2^+$ with an accuracy of 2.3 MHz, limited by the bandwidth of the pulsed laser THz radiation used in the experiments. The spectral resolution was, however, not sufficient to fully resolve the fine structure of the Rydberg series so that the spin-rotation interval of the $N^+=2$ ionic level could not be determined. The possibility to resolve the fine structure of the $n$f$N^+=2$ series and to determine the spin-rotational interval of the $N^+=2$ ionic level was later demonstrated in the thesis of Beyer \cite{beyer18e}. In the present work, the use of frequency-comb-calibrated continuous-wave single-mode laser radiation enabled a significant improvement of the precision and accuracy of the measurements and the recording of fully resolved fine-structure patterns over a broad range of principal quantum numbers for the $n$f$N^+=2$, 4 and 6 Rydberg series. 

Finally, our study aimed at determining the rotational and spin-rotational structure in the ground vibronic state of para-H$_2^+$ by combining precision Rydberg spectroscopy and MQDT, focusing on the spin-rotation fine structure that could not be determined in Ref.~\cite{haase15a}. H$_2^+$ does not have electric-dipole-allowed rovibrational transitions and Rydberg-series extrapolation is an attractive approach to determining its level structure. While spin-rotation intervals in higher vibrational levels of H$_2^+$ have been measured very precisely (precision in the kHz range) by Jefferts \cite{jefferts69a} with magnetic dipole transitions, no experimental values for the spin-rotation intervals in the ground vibronic state have been reported to date.

\begin{figure*}
	{\includegraphics[trim=0cm 0.0cm 0cm 0cm, clip=false, width=0.9\linewidth]{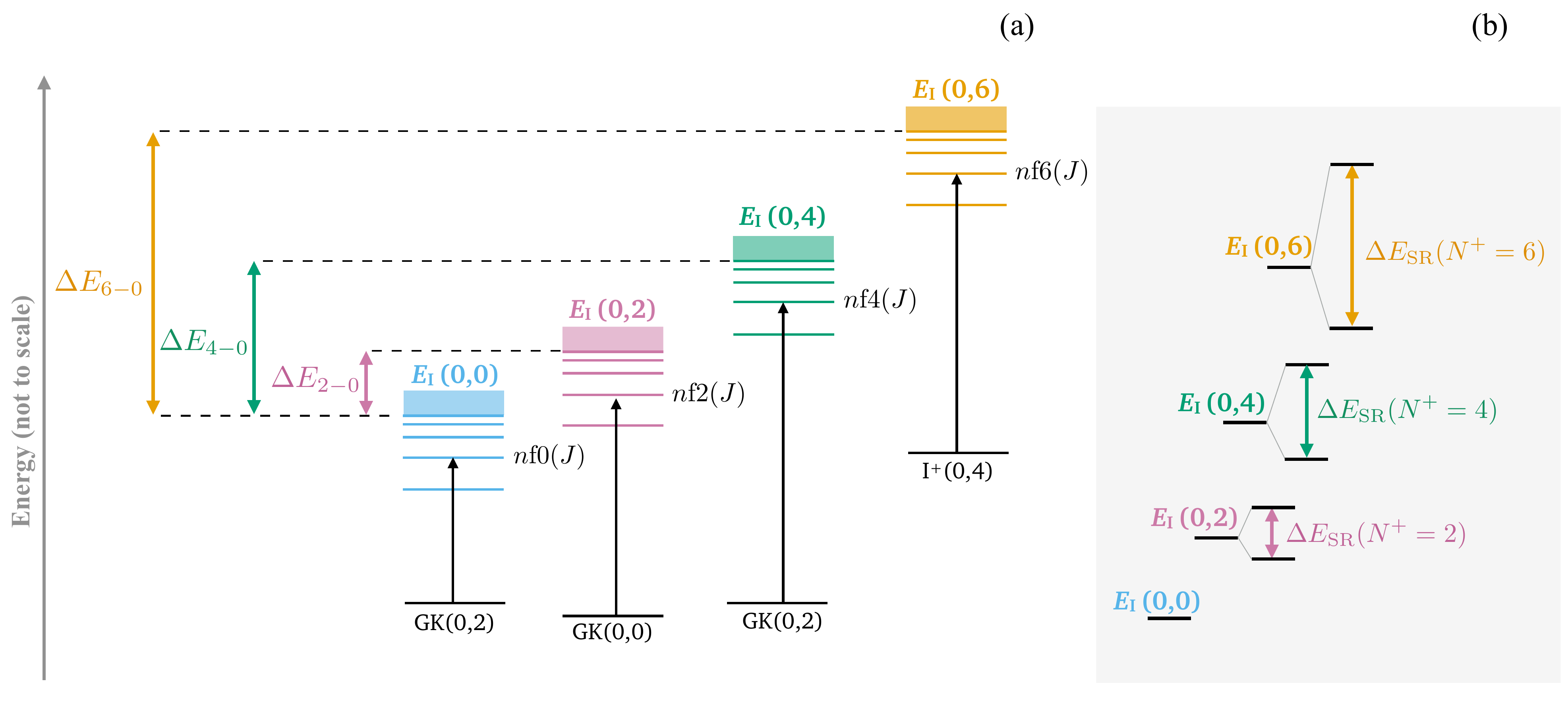}}
      \caption{(a) Schematic energy-level diagram (not to scale) showing the transitions to $n$f Rydberg states of para-H$_2$ used to determine the $\Delta E_{2-0}$, $\Delta E_{4-0}$, and $\Delta E_{6-0}$ rotational intervals of the X$^+$~$^2\Sigma_g^+\,(v^+=0)$ ground state of para-H$_2^+$ by Rydberg series extrapolation. (b) Detailed view (not to scale) including the spin-rotation splittings of the $N^+=0, 2, 4$ and 6 levels of H$_2^+$.}  
              \label{fig:Ediagram}	
\end{figure*}

In the para-H$_2^+$ ion core, the spin-rotation interaction leads to splittings of the ionic energy levels (see panel (b) of Fig. \ref{fig:Ediagram}) 
\begin{equation}\label{eq:spinrot_int}
      \Delta E_{\textrm{SR}}(N^+) = hc\gamma ^{N^+}_{\textrm{SR}}(N^+ + \frac{1}{2}), 
\end{equation}
that rapidly increase with the rotational quantum number $N^+$ of the ion core. In Equation~\eqref{eq:spinrot_int}, $\gamma ^{N^+}_{\textrm{SR}}$  is the $N^+$-dependent spin-rotation coupling constant which, at low $N^+$ values, can be well described by retaining only the first-order centrifugal-distortion contribution (see, e.g., Refs.~\cite{kristensen90a,jansen18b}):
\begin{align}
      \gamma ^{N^+}_{\textrm{SR}} = \gamma ^{(0)} + \gamma ^{(1)} N^+ (N^+ + 1).
      \label{eq:gamma_dep}
\end{align}

The spin-rotation interaction in the ion core also significantly affects the fine structure of high-$n$ Rydberg states and is the driving force, in para-H$_2$, of the uncoupling of the Rydberg electron as $n$ increases, as explained above.

Fig. \ref{fig:Ediagram} illustrates the principle of our measurements. Transitions to Rydberg states belonging to series converging to the $v^+=0, N^+= 2, 4$ and 6 levels of the X$^+$ $^{2} \Sigma_g^+$ state of H$_2^+$ are measured from the GK $^1\Sigma_g^+(v=0,N=0,2)$ and I$^+$ $^1\Pi_g^+(v=0,N=4)$ intermediate states of H$_2$, which are themselves populated from the X $^1\Sigma_g^+(v=0,N=0,2)$ state by resonant two-photon excitation via the B $^1\Sigma_u^+(v=4,N=1,3)$ states. The resonant nature of the excitation scheme enables a clean selection of para-H$_2$ and avoids the generation of ions by multiphoton ionization.
In the remainder of this article, the Rydberg states are labeled $n \ell N^+ (J)$, where $n$ is the principal quantum number, $\ell$ is the quantum number associated with the orbital angular momentum of the Rydberg electron, $N^+$ is the rotational angular momentum quantum number of the ion core, and $J$ is the quantum number for the total angular momentum in para-H$_2$ (total nuclear spin $I=0$). 

\section{Experimental Procedure} \label{sec:exp}

The Rydberg states are accessed using the resonant three-photon excitation sequence
\begin{align}
    \text{X}\,^1\Sigma_g^+(0,N''=0,2)  \xrightarrow{\mathrm{VUV}} &~\text{B}\,^1\Sigma_u^+(4,N'=1,3) \notag\\
     \xrightarrow{\mathrm{VIS}} &
     \begin{aligned}[t]
         &~\text{GK}\,^1\Sigma_g^+\, (0,N=0,2) \\
         &~\text{or I}^+ \,^1\Pi_g^+\,(0,N=4)
     \end{aligned} \label{eq:excischeme} \\
     \xrightarrow{\mathrm{NIR}} &~n\mathrm{f} \,\,[\text{H}_2^+\,\text{X}{^{+}}\,^{2} \Sigma_g^+ (0,N^+=2,4,6) ]\,.\notag
\end{align}
Rydberg states converging to the $N^+=2$ and $N^+=4$ rotational levels of H$_2^+$ are excited from the X $^1\Sigma_g^+(0,0)$ ground state via the B $^1\Sigma_u^+(4,1)$ and GK $^1\Sigma_g^+(0,0)$ ($\tau \approx$ 79 ns \cite{hoelsch18a}) intermediate states and via the B $^1\Sigma_u^+(4,1)$ and GK $^1\Sigma_g^+(0,2)$ ($\tau \approx$ 58 ns \cite{hoelsch18a}) intermediate states, respectively. For Rydberg states converging to the $N^+=6$ threshold of H$_2^+$, the X $^1\Sigma_g^+(0,2)$, B $^1\Sigma_u^+(4,3)$ and I$^+$ $^1\Sigma_g^+(0,4)$ ($\tau \approx$ 15-20 ns, determined using a pump-probe scheme analogous to that used in Ref. \cite{hoelsch18a}) states are used. In this case, the I$^+$(0,4) state is chosen instead of the GK(0,4) state because of its much stronger transitions to $n$f6 Rydberg states. Accessing the I$^+$(0,4) state in a two-photon process requires using the rotationally excited X $^1\Sigma_g^+(0,2)$ state, which is weakly populated in a supersonic expansion, as a starting level. When the GK(0,0) state is used as intermediate state, the excitation to $n$f Rydberg states is restricted to states with $J=1$. Transitions from the GK(0,2) and I$^+$(0,4) levels provide access to Rydberg states with total angular momentum quantum number $J=1-3$ and $J=3-5$, respectively. However, the only strong and narrow spectral lines correspond to $n$f4$(J=1)$ and $n$f6$(J=3)$ Rydberg states. 
\par
The experimental apparatus and measurement procedure are similar to those described in detail in Ref.~\cite{beyer18a}. They are therefore only briefly summarized here. A pulsed (repetition rate 25 Hz) skimmed supersonic expansion of pure H$_2$ is emitted from a valve held at 60 K for the experiments using the ground rovibrational X$(0,0)$ state as a starting level. Because cooling to 60 K depletes the occupation probability of the X$(0,2)$ level to below 0.1\%, the valve is held at room temperature in the experiments accessing $n$f6$(J)$ Rydberg states. Nonetheless, the partial cooling of the rotational degrees of freedom in the supersonic expansion strongly reduces the density of X$(0,2)$ rotationally excited molecules in the photoexcitation region. Consequently, the spectra of the $n$f6 Rydberg states have a much lower signal-to-noise ratios than those of $n$f2 and 4 Rydberg states.

\par
The intermediate GK and I$^+$ states are accessed using pulsed vacuum-ultraviolet (VUV, $\lambda \approx 105$ nm) and visible (VIS, $\lambda \approx 613$ nm or 580 nm) laser radiation, as detailed in Ref.~\cite{beyer18a}. 
The Rydberg states are excited from the selected GK or I$^+$ intermediate levels using a single-mode continuous-wave (cw) Ti:Sa laser ($\lambda$ between 750 and 780 nm) with a bandwidth of 100 kHz. To achieve a high absolute frequency accuracy ($\Delta \nu / \nu$ in the order of $2\cdot 10^{-11}$), the cw-laser frequency is stabilized to a frequency comb that is referenced to a Rb oscillator disciplined by a Global-Positioning-System receiver. To minimize systematic uncertainties arising from first-order Doppler shifts, the cw-laser beam intersects the molecular beam at near right angles and the laser beam is retroreflected by a mirror placed beyond the photoexcitation region. A small deviation angle $\alpha$ from 90$^\circ$ is intentionally maintained, resulting in each spectral line having two Doppler components with first-order Doppler shifts $\pm (v\sin{\alpha}/c) \, \nu_{\textrm{laser}}$ of equal magnitude but opposite sign. The first-order Doppler-free transition frequencies are obtained as averages of the fitted central positions of the two Doppler components. To ensure precise retroflection and wavefront retracing, the laser beam is focused at the reflection mirror with a lens system of large effective focal length and the incoming and retroreflected laser beams are aligned at a distance of $\approx 12$~m from the mirror, resulting in a retroreflection-angle control of better than 100~$\mu$rad \cite{beyer18a}.
\par
To reduce systematic uncertainties from dc-Stark shifts, residual stray electric fields are compensated in three dimensions by applying well-controlled dc electric fields in the photoexcitation region and minimizing the Stark shifts, as explained in Refs.~\cite{osterwalder99a,beyer18a}. Pulsed electric fields are used to extract the H$_2^+$ ions generated by autoionization of the Rydberg states located above the X$^+$(0,0) ionization threshold towards a microchannel-plate (MCP) detector. These fields also field ionize long-lived Rydberg states, including those located below the X$^+$(0,0) ionization threshold, and extract the resulting ions for detection. The spectra are recorded by monitoring the H$_2^+$-ion current at the MCP detector as a function of the frequency of the cw laser.

\section{Qualitative Aspects of the Energy-Level Structure and Experimental Results} \label{sec:results}
\begin{figure*}
	{\includegraphics[trim=0cm 0.0cm 0cm 0cm, clip=false, width=0.9\linewidth]{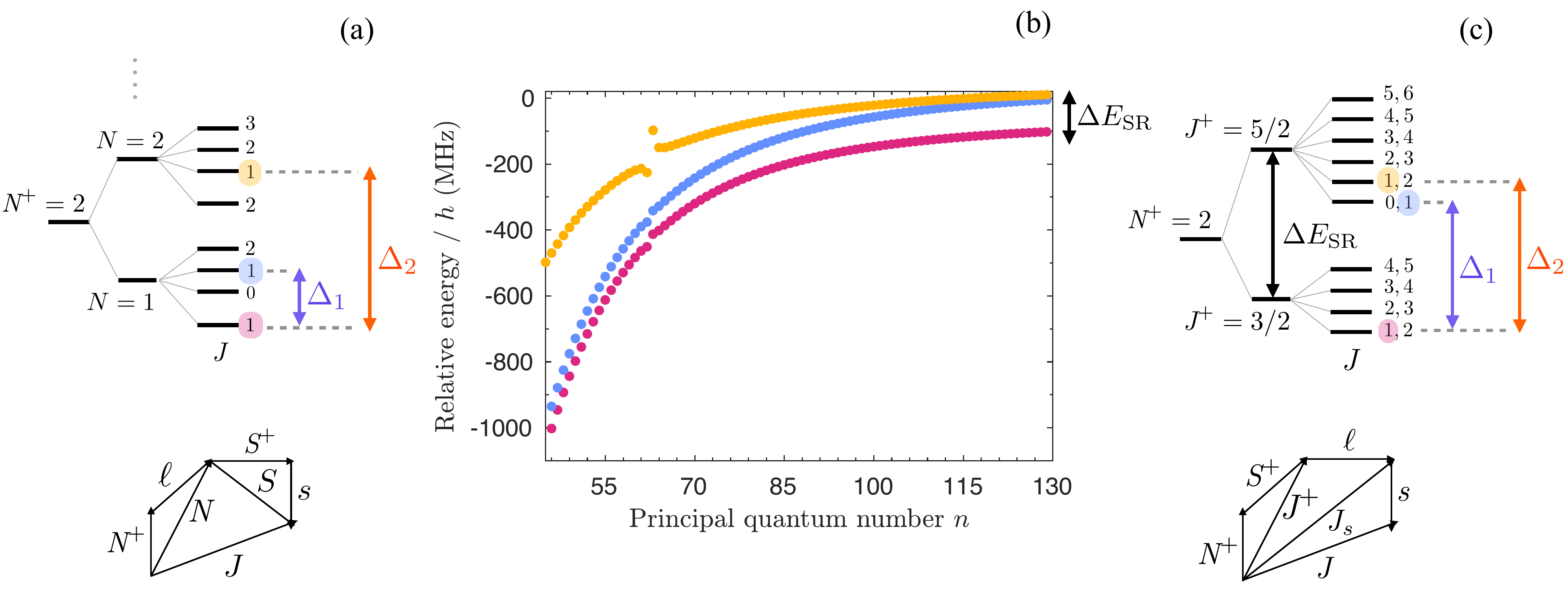}}
      \caption{(a) and (c) Qualitative energy-level diagrams and angular-momentum coupling schemes corresponding to f Rydberg states of H$_2$ with an $N^+=2$ ion core, for low $n$ ($n<50$) and high $n$ ($n>130$), respectively. $\Delta_1$ (purple) and $\Delta_2$ (orange) designate the spacings between the three $J=1$ fine-structure components. (b) Evolution, from low to high $n$ values, of the positions of the three $n$f2$(J=1)$ Rydberg states relative to the zero-quantum-defect positions $-cR_{\textrm{H}_2}$/$n^2$, obtained in an MQDT calculation including electron spins. The magenta, blue, and yellow dots designate these three states in order of increasing energy, as displayed in panels (a) and (c). The double-headed arrow indicates the spin-rotation splitting of the $N^+=2$ ion core, shown on an expanded scale in panel (c).}  
     \label{fig:levelstr}	
\end{figure*}

The energy-level structure of $n$f Rydberg states belonging to series converging to excited rotational levels of the X$^+$ $^2\Sigma_g^+$ ground electronic state of para-H$_2^+$ is depicted schematically in Fig.~\ref{fig:levelstr}. Nuclear spins do not contribute to the structure of even-$N^+$, low-$v^+$ levels of the X$^+$ $^2\Sigma_g^+$ electronic ground state of H$_2^+$ because nuclear-spin symmetry is conserved in these states and even-$N^+$ states have $I=0$ total nuclear spin. The level structure is determined by the relative strengths of the interactions that couple the rotational angular momentum $\vec{N^+}$ of the ion core, the core-electron spin $\vec{S^+}$, and the orbital angular momentum $\vec{\ell}$ and the spin $\vec{s}$ of the Rydberg electron.
 These interactions are (i) the exchange interaction between the Rydberg and ion-core electrons, which causes a splitting between singlet $S=0$ and triplet $S=1$ states; (ii) the long-range charge-quadrupole and polarization electrostatic interactions that couple $\vec{\ell}$ and $\vec{N^+}$; (iii) the spin-orbit interactions that couple $\vec{\ell}$ with $\vec{s}$ and $\vec{S^+}$; and (iv) the spin-rotation interaction and higher-order spin-orbit interactions that couple $\vec{N^+}$ and $\vec{S^+}$ in the ion core. 

 The strengths of interactions (i)-(iii) are proportional to the Rydberg-electron density in the H$_2^+$-core region and thus scale with the principal quantum number $n$ as $n^{-3}$. The spin-orbit interactions (iii) are several orders of magnitude weaker than interactions (i) and (ii). They are so weak in the $n$f Rydberg states with $n>50$ of interest here that their effects on the level structure are not observable within the limits of our experimental resolution and sensitivity. They are therefore neglected in the following. The spin-rotation interaction (iv) and the ion-core rotational spacings are properties of the ion core and therefore $n$-independent. Consequently, interactions (i) and (ii) are expected to determine the main patterns of the level fine structure at low-$n$ values, whereas interaction (iv) is expected to determine its main patterns at high $n$ values. 

 Panels (a) and (c) of Fig. \ref{fig:levelstr} display the angular-momentum coupling schemes corresponding to these two, low-$n$ and high-$n$, limiting situations, as well as the resulting qualitative energy-level diagrams for $N^+=2$. For Rydberg states with $n$ up to about 50 [panel (a)], the electrostatic coupling between $\vec{\ell}$ and $\vec{N^+}$ dominates, such that $N$  ($\vec{N}$ =  $\vec{\ell}$ + $\vec{N^+}$) is a good quantum number and the levels form groups that can be labeled by the value of $N$ ($N=1-5$ for $n$f2 series). Within each group, the exchange interaction further splits the structure according to the spin multiplicity, the singlet $J=N$, $S=0$ level being located below the triplet $J=N,N\pm 1$, $S=1$ levels. Singlet-triplet mixing among the two close-lying components of the same $N$ values is already significant at this $n$ value. At $n \approx 130$, the spin-rotation interaction in the ion core (interaction (iv)) becomes dominant and decouples $\vec{\ell}$ from $\vec{N^+}$ and $\vec{s}$ from $\vec{S^+}$. $J^+$ ($\vec{J^+}$ = $\vec{N^+}$ + $\vec{S^+}$) becomes the pattern-determining  quantum number, giving rise to two groups of levels with $J^+=3/2$ and 5/2 separated energetically by the spin-rotation splitting of the $N^+=2$ ion-core level [panel (c)]. Within these groups, the levels are further split by the electrostatic interaction (ii) which couples $J^+$ with $\vec{\ell}$ to form $\vec{J}_s$. The Rydberg-electron spin is almost fully decoupled in this situation and the levels occur as pairs of near-degenerate states with $J=J_s\pm 1/2$.
 
In the treatment of the level structure of H$_2$ Rydberg states by MQDT, interactions (i)-(iv) and the evolution of the level structure from the low-$n$ to the high-$n$ case are accounted for by the values of the quantum-defect parameters and the ion-core level structure used in the calculations and which can be determined from fits to experimentally observed level positions or in ab initio calculations \cite{jungen11a,sprecher14x}, as explained further in Section~\ref{sec:mqdt}.  As illustration of this evolution, Fig. \ref{fig:levelstr}(b) shows the level structure of $n$f2($J=1$) Rydberg states obtained in an MQDT calculation including electron spins based on quantum-defect parameters extracted from the recent high-level ab initio calculations of low-$n$ Rydberg states \cite{silkowski21a,silkowski22a,silkowski23a,silkowski24a}, as explained in Section~\ref{subsec:mqdt_theory}. The calculated energies of the three $J=1$ fine-structure components are shown as magenta, blue, and yellow dots, in order of increasing energies at each value of $n$. The intervals between these three components change with the principal quantum number $n$ because the hierarchy of angular-momentum coupling gradually changes between the situations depicted schematically in panels (a) and (c).  At $n=50$, the largest energy separation is between the two almost degenerate states with $N=1$ (magenta and blue dots) and the state with $N=2$ (yellow dots).  As $n$ increases, the two $N=1$ states split, and the energy of the middle component (blue dots) starts approaching that of the highest-energy state (yellow dots). At $n \approx 130$, the pattern consists of one single level at low energies, which can be associated with the lower ($J^+=3/2$) ion-core spin-rotational component, and a doublet at high energies corresponding to the upper ($J^+=5/2$) ion-core spin-rotational component. This change in behavior is the result of the $n$-independent spin-rotation interaction (iv) becoming dominant over interactions (i) and (ii), which both scale as $n^{-3}$, as explained above. 
As $n \rightarrow \infty$, the splitting between the states with $J^+=\frac{3}{2}$ and with $J^+=\frac{5}{2}$ converges to the spin-rotation splitting of the $N^+=2$ H$_2^+$ ion-core level, which is indicated in panel (b) by the double-headed arrow. The local perturbation in the energy-level structure centered around the 62f2$(J=1)$ states in panel (b) is the effect of a rotational-channel interaction with the 16f4$(J=1)$ states.

The behavior of the three components for each of the $n$f4$(J=1)$ and $n$f6$(J=3)$ states is similar to that just discussed for the $n$f2$(J=1)$ states and is illustrated in Figs. S1 and S2 of the supplemental material, respectively. The main difference is that the spin-rotation splitting of the ion core increases with $N^+$ [see Eq.~\eqref{eq:spinrot_int}], so that the limiting situation illustrated by Fig. \ref{fig:levelstr}(c) is reached at lower $n$ values. The energy-level structures of $n$f Rydberg states with different values of $J$ show similar patterns and can be rationalized using the same angular-momentum-coupling arguments. Transitions to these Rydberg states are, however, weak and significantly broadened by autoionization, and therefore not analyzed further.

\begin{figure*}
	{\includegraphics[trim=0cm 0.0cm 0cm 0cm, clip=false, width=0.72\linewidth]{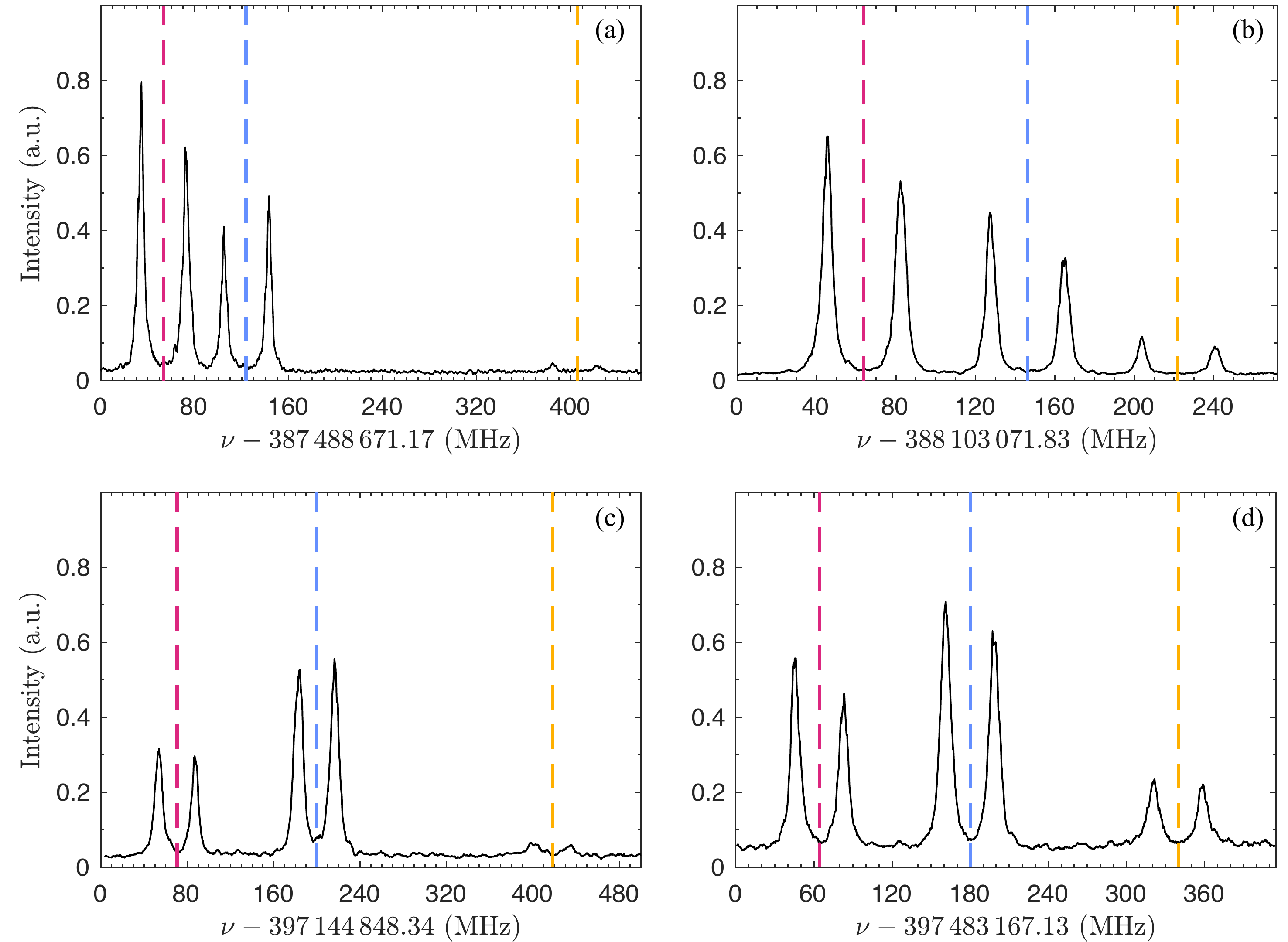}}
      \caption{Spectra of the 54f2$(J=1)$ $\leftarrow$ GK(0,0) (a), 80f2$(J=1)$ $\leftarrow$ GK(0,0) (b), 46f4$(J=1)$ $\leftarrow$ GK(0,2) (c), and 52f4$(J=1)$ $\leftarrow$ GK(0,2) (d) transitions. The vertical dashed lines indicate the fitted Doppler-free transition frequencies for each of the three fine structure components with $N^+=2$, $J$ = 1 ($N^+=4$, $J$ = 1). The color coding of the vertical lines corresponds to that shown in Fig. \ref{fig:levelstr}. }  
              \label{fig:Np2_Np4_spectra}	
\end{figure*}
As typical examples of the spectral patterns just discussed, Fig. \ref{fig:Np2_Np4_spectra} depicts spectra of $n$f2$(J=1)$ Rydberg states at $n=54$ (a) and $n=80$ (b) recorded from the GK(0,0) intermediate level, and $n$f4$(J=1)$ Rydberg states at $n=46$ (c) and $n=52$ (d) states recorded from the GK(0,2) level. These spectra consist of three pairs of lines, each corresponding to the two Doppler components of a transition to a specific fine-structure level. The colored vertical lines, with color code corresponding to Fig. \ref{fig:levelstr}, indicate the first-order-Doppler-free positions. These positions are determined from fitting Voigt line profiles to the spectral lines. Their full widths at half maximum are all in the range between 6 and 9 MHz and are only slightly larger than the widths of 5 MHz obtained in measurements of $n$f series converging to the $N^+=0$ level of H$_2^+$ \cite{beyer18a}. This line broadening is attributed to rotational autoionization because all final states of these transitions have rotationally excited H$_2^+$ ion cores and lie above the lowest ionization threshold of H$_2$, corresponding to the X$^+(v^+=0,N^+=0)$ ground state of H$_2^+$.

\begin{figure}[h]\centering
	{\includegraphics[trim=0cm 0.0cm 0cm 0cm, clip=false, width=0.82\linewidth]{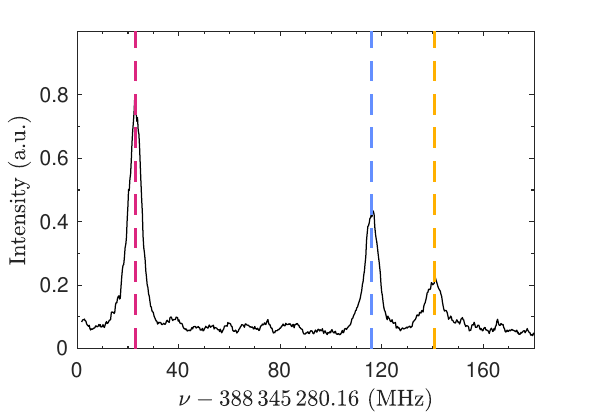}}
      \caption{Spectrum of the 110f2$(J=1)$ $\leftarrow$ GK(0,0) transition measured without the retroreflected beam. The vertical dashed lines indicate the fitted transition frequencies for each of the three fine structure components with $N^+=2$, $J$ = 1, corresponding to the measured Doppler component. The color coding of the vertical lines corresponds to that shown in Fig. \ref{fig:levelstr}.}  
              \label{fig:Np2_n110_spectrum}	
\end{figure}

In Fig.~\ref{fig:Np2_Np4_spectra}(a), corresponding to the low-$n$ coupling situation, the transitions to the lowest two fine-structure components of the 54f2$(J=1)$ Rydberg state are strong and the transition to the third component is much weaker. The lowest two components are the two $N=1$ states of mixed singlet-triplet character [see Fig. \ref{fig:levelstr}(a)] that can be reached in a single-photon transition from the GK $^1\Sigma_g^+$ state with $N=0$. In the approximation of $N$ being a good quantum number, the third component with $J=1$ is a triplet state with $N=2$, which is not optically accessible from the $N=0$ starting level. Weak lines are nevertheless observed for this transition, because at $n=54$ the spin-rotation interaction in the ion-core already slightly mixes states of different $N$ values. At $n=80$ [panel (b)], corresponding to an intermediate situation between the low- and high-$n$ coupling situations, the transitions to all three fine-structure components exhibit comparable intensities, because the spin-rotation interaction in the ion core already significantly mixes singlet and triplet states of different $N$ values (see Fig. \ref{fig:levelstr}(c)). The spectrum of the 110f2(1) states displayed in Fig.~\ref{fig:Np2_n110_spectrum} was recorded without the back-reflected beam and consists of one Doppler component per transition. It corresponds to the high-$n$ coupling situation. 
The spectral pattern exhibits a main splitting, given by the ion-core spin-rotation interval, between the lower component and the upper two close-lying components, as expected from Fig.~\ref{fig:levelstr}(b). 

The spectra of the 46f4$(J=1)$ and 52f4$(J=1)$ states displayed in Figs. \ref{fig:Np2_Np4_spectra}(c) and (d), respectively,  display a similar behavior as the $n$f2$(J=1)$ states, with the difference that the "forbidden" transition to the highest ($N=2$) fine-structure component is already significant at $n$ values below $54$. This difference originates from the larger spin-rotational splitting of the $N^+=4$ ion core level, as already discussed above.

\begin{figure}[h]\centering
	{\includegraphics[trim=0cm 0.0cm 0cm 0cm, clip=false, width=0.82\linewidth]{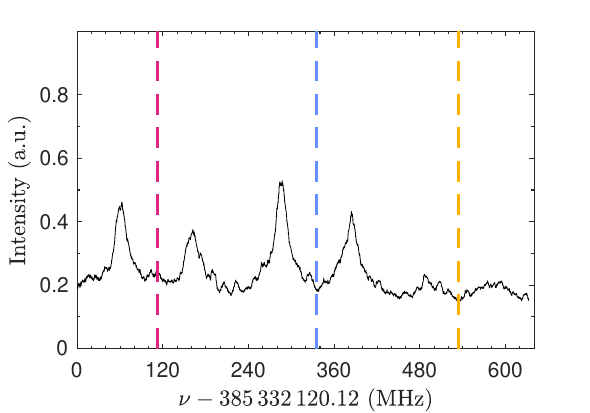}}
      \caption{Spectrum of the 46f6$(J=3)$ $\leftarrow$ I$^+$(0,4) transition. The vertical dashed lines indicate the fitted Doppler-free transition frequencies for each of the three fine structure components with $N^+=6$, $J$ = 3. The color coding of the vertical lines corresponds to that shown  in Fig. \ref{fig:levelstr}.}  
              \label{fig:Np6_spectra}	
\end{figure}
The spectrum of the 46f6$(J=3)$ $\leftarrow$ I$^+$(0,4) transition, displayed in Fig. \ref{fig:Np6_spectra}, also consists of three pairs of lines with Doppler-free transition frequencies indicated by colored dashed vertical lines using the same color coding as in Figs.~\ref{fig:levelstr} and \ref{fig:Np2_Np4_spectra}.  Compared to the spectra of Rydberg states with $n$f2 and $n$f4 Rydberg states discussed above, the full widths at half maximum of about 25~MHz are significantly larger and the signal-to-noise ratio is strongly reduced. The increase in linewidths is attributed to the shorter lifetime ($\tau \approx$ 15-20 ns) of the I$^+$(0,4) state, corresponding to a Lorentzian contribution of $\approx$ 8-11 MHz, and to the lower quality and broader transverse-velocity distribution of the supersonic expansion caused by the valve being held at room temperature instead of 60 K to enhance the population of the $J=2$ ground state rotational level, as explained in Section \ref{sec:exp}. Moreover, the lifetime of the I$^+$(0,4) state is comparable to the pulse length of the VIS laser ($\approx$ 8 ns) used to access the Rydberg states starting from the I$^+$ state. Therefore, the excitation to Rydberg states takes place to a significant extent while the pulsed VUV and VIS lasers are on, which can result in ac-Stark broadening and shifts (see below).  

The frequencies of numerous transitions to $n$f2($J=1$), $n$f4($J=1$), and $n$f6($J=3$) Rydberg states converging to X$^+$ $^2\Sigma_g^+ (v^+=0,N^+=2,4,6)$ rotational levels of H$_2^+$ were measured in the range of principal quantum number $n$ between 28 and 115 and the results are summarized in panels (a)-(c) of Fig. \ref{fig:rotint_resi} below. These frequencies are the input required to determine the rotational and spin-rotational fine structure of the H$_2^+$ ion core in the MQDT fits and the quantification of their uncertainties is of key importance in the present work. Each frequency was obtained as the average of several measurements carried out after full realignment of the optical setup. In this way,
the systematic uncertainties caused by residual first-order Doppler shifts average out and can thus be treated as statistical.

\begin{table}[h]
\centering
\caption{Error budget and frequency corrections with the example of the determination of the transition frequency to the lowest-energy fine-structure component of the 
54f2$(J=1)$ state from the GK(0,0) intermediate level (measured frequency $\tilde\nu_{\textrm{laser}}$ =12\,925.232\,572 cm$^{-1}$) from a set of independent measurements after complete realignment of the optical setup. All values and uncertainties are reported in kHz.}
\label{tab:errtr}
\begin{tabular}{llll} 
\toprule
 & $\Delta \nu$  &  \ \ $\sigma _\textrm{stat}$  & \ \ $\sigma _\textrm{sys}$ \\
\midrule
Lineshape fits and  & & \ \ & \\ 
residual 1$^ {\textrm{st}}$-order Doppler shift & & \ \ 250 & \\ 

2$^ {\textrm{nd}}$-order Doppler shift & \ \ +2 & &  \ \ 0.5 \\
dc Stark shift &  & & \ \ 25 \\
ac Stark shift &  & & \ \ $<$5 \\
Zeeman shift &  &  & \ \ $<$10 \\
Pressure shift &  &  & \ \ $<$1 \\
Photon-recoil shift & $-$165  & & \\
\bottomrule
\end{tabular} 
\end{table}
The error budget for a typical transition frequency is presented in Table \ref{tab:errtr}. The main contribution to the overall uncertainty (250 kHz) is statistical and originates, in almost equal parts, from the intrinsic uncertainties in the determination of the line centers through fits of the line shapes with Voigt profiles and from the distribution of the residual first-order Doppler shifts. The systematic uncertainties corresponding to the 54f2$(J=1)$ $\leftarrow$ GK(0,0) transition are also listed in Table \ref{tab:errtr}. They were determined in series of measurements in which the experimental conditions (laser intensities, beam velocity, valve stagnation pressure) were systematically modified.  In the case of transitions to $n$f2($J=1$) and $n$f4($J=1$) states, the systematic uncertainties are much smaller than the statistical ones and do not exceed 25 kHz. As discussed above, the measurements of the transition frequencies to the $n$f6($J=3$) states are possibly affected by ac-Stark shifts caused by the pulsed laser radiation during the short lifetime of the I$^+$ state.  It was not possible to fully quantify these shifts in our experiments, but we estimate the maximum possible systematic uncertainty to be 5 MHz and we include this contribution when determining the systematic uncertainty of the X$^+ (0,6) - $I$^+(0,4)$ interval (see Table \ref{tab:rotint} in Section~\ref{subsec:mqdt_fits}). All transitions frequencies determined in this work, after correcting for their precisely known second-order Doppler shifts ($+ 2$~kHz) and the photon-recoil shifts ($-165$~kHz), are listed in the supplemental material. 

The energy intervals between the GK(0,0), GK(0,2) and I$^+$(0,4) states needed to derive a common relative energy scale for the $n$f$N^+(J)$ Rydberg states are listed in Table~\ref{tab:rotint}. The relative position of the GK(0,0) and GK(0,2) is known with sufficient accuracy from earlier work \cite{beyer18a}. The relative position of the I$^+(0,4)$ and GK(0,2) states was determined by measuring the transition frequencies from each of these two states to the 56f4$(J=3)$ Rydberg state. This state lies in close proximity of, and interacts with, the 13f6$(J=3)$ Rydberg state, and the resulting mixed $N^+=4$ and 6 character makes this state accessible in one-photon transitions from both the I$^+(0,4)$ and the GK(0,2) levels. The low population of the X(0,2) level of H$_2$ used to access the ${\rm I}^+(0,4)$ state, however, limited the signal-to-noise ratio of the spectrum of the 56f4$(J=3)\leftarrow {\rm I}^+(0,4)$ transition. Consequently, the interval between the I$^+(0,4)$ and GK(0,2) levels could only be determined with an uncertainty of 2~MHz.

\section{MQDT Analysis} 
\label{sec:mqdt}
Over the years, MQDT, which was initially introduced to explain autoionization and perturbations caused by rotational channel interactions in the Rydberg spectrum of H$_2$ \cite{fano70a,herzberg72a}, has been continuously refined and extended to account for the structure and dynamics of molecular Rydberg states and to determine accurate molecular ionization energies \cite{atabek74a,jungen77a,greene85a,du86a,jungen90a,jungen90b,huber90a,huber94a,jungen97a,jungen97b,jungen98a,osterwalder04a,jungen11a,kay11a,sprecher14a,sprecher14x,sommavilla16a}. In the case of molecular hydrogen, a consistent set of quantum-defect parameters, derived from ab initio calculations of low-lying Rydberg states and fits to experimentally observed line positions, has been established that can describe the $n$p and and $n$f series with high accuracy, including subtle effects related to the fine and hyperfine structures \cite{osterwalder04a,woerner07a,hoelsch19a}. 

The recent possibility to calculate the potential energy functions of the low-lying Rydberg states of H$_2$ up to $n\approx 10$ ab initio with almost arbitrary accuracy \cite{silkowski21a,silkowski22a,silkowski23a} opens a route to the determination of full sets of quantum-defect parameters, including their energy dependence, following procedures established by Jungen and his coworkers \cite{ross87a,ross94a,ross94b,ross94c,osterwalder04a,sprecher14x}. To analyze the data on $n$f Rydberg states presented in the previous section, we use a new set of quantum-defect parameters determined from recent high-level calculations of the low-lying Rydberg states of H$_2$ \cite{silkowski21a,silkowski22a,silkowski23a,silkowski24a}. Given the exceptional accuracy of these calculations, with estimated uncertainties in the order of 10 MHz at $n=8$, and the $n^{-3}$ scaling of the errors in the positions calculated from quantum defects, these quantum-defect parameters are expected to be sufficiently accurate to describe the positions of high-lying Rydberg states (with 40 kHz accuracy at $n$=50), except in the vicinity of perturbations involving low-lying Rydberg states with vibrationally excited ion cores, as detailed in Section~\ref{subsec:mqdt_fits}.

In the following, we briefly summarize the MQDT formalism used to calculate the positions of high-lying $n$f Rydberg states for comparison with the experimental results and we present the main features of the new quantum-defect parameters. All calculations were performed using the MQDT-programs developed over several decades in the group of Christian Jungen.

\subsection{MQDT Formalism and Quantum Defects Parameters}
\label{subsec:mqdt_theory}
The term value $T_n$ of a Rydberg state with principal quantum number $n$ is given by 
\begin{equation}\label{eq:Rydeq1}
    T_n=\frac{E^{+}\left(v^{+}, N^{+}, J^{+}\right)}{h c}-\frac{\mathcal{R}_{H_2}}{(n-\bar{\mu})^2},
\end{equation}
where $E^+$ is the ionization energy and $\bar{\mu}$ the effective quantum-defect of this specific Rydberg channel.   

For $n\ell$ Rydberg states, the MQDT formalism allows the determination of the effective quantum-defect in Eq.~\eqref{eq:Rydeq1} for all possible combinations of vibrational and rotational excitation in the ion core and for all possible combinations of the spin of the core and the Rydberg electron using the concept of a \emph{frame transformation} \cite{fano70a,jungen77a,jungen98a}. This concept is based on the idea, that the scattering Rydberg electron and the ion core can be described in Hund's case (b), while the Rydberg equation in \eqref{eq:Rydeq1} assumes Hund's case (d) coupling.

In the specific case of a Rydberg electron with $\ell=3$ and $s=1/2$ and the ion core in the $^2\Sigma_g^+$ state, $\Lambda = 0, 1, 2$ and 3 for $S= 0,1$ ($\vec{S}$ =  $\vec{S^+}$ + $\vec{s}$). This results in 8 eigenquantum defects $\mu_\text{f}^{(S\Lambda)}$ for the description of f Rydberg states in para-H$_2$ (note: (i) we neglect the nuclear spin because $I=0$ in para-H$_2$) and (ii) the quantum defects have no dependence on $\Omega$ \cite{osterwalder04a}). 

The quantum defects depend on the internuclear distance $R$ and the energy of the Rydberg electron $\varepsilon$. Following the treatment of the p Rydberg states in Ref.~\cite{sprecher14a}, they can be represented as 24 functions of the internuclear distance: 
\begin{equation}\label{eq:qdset}
    \mu_\text{f}^{(S\Lambda)}(R,\varepsilon=0), \frac{d\, \mu_\text{f}^{(S\Lambda)}(R,\varepsilon=0)}{d\, \varepsilon}, \frac{d^2\, \mu_\text{f}^{(S\Lambda)}(R,\varepsilon=0)}{d\, \varepsilon^2}.
\end{equation}
The eigenquantum defects and their energy derivatives are evaluated at the ionization threshold, i.e., for zero binding energy of the Rydberg electron.

In Ref.~\cite{osterwalder04a}, eigenquantum-defect functions for f Rydberg states were extracted from a polarization model, which does not distinguish between singlet and triplet ion-core--Rydberg-electron scattering. In addition, the polarization model assumes a multipole expansion for the ion core and diverges for larger $R$ values. However, the model enables the calculation over a large range of principal quantum numbers, which results in a better extrapolation of the quantum defects to $\varepsilon=0$.

While the $R$-dependence of the eigenquantum defects predominantly affects the positions of vibrational interlopers, the $\Lambda$- and $S$-dependence has an impact on the spin-rotational structure of the Rydberg states, which represents one of the main topics of this work. 

The extraction of the quantum defects from the available ab intio data follows the procedure described in Ref.~\cite{sprecher14a} and a more detailed account will be given in a future publication~\cite{silkowski24a}. In short, the PES are converted to $\mu_\ell^{(S\Lambda)}(R,n)$ , which enables the identification of the states with f character and distinguishes them from the p (larger quantum defect) and h (smaller quantum defect) states. For each value of the internuclear distance $R$, the quantum defects for a given $(S\Lambda)$ pair are extrapolated to $n=\infty$ using a second-order polynomial, corresponding to the zero-energy quantum defect $\mu_\text{f}^{(S\Lambda)}(R,\varepsilon=0)$. For $\varepsilon=0$, the $\mu$-quantum defects correspond to the $\eta$-quantum defects, introduced by Ham \cite{ham55a} to avoid unphysical solutions when applying the MQDT quantization condition. The energy dependence of the $\eta$-defects is encoded similarly to the one in Eq.~\eqref{eq:qdset} and is obtained in a least-squares adjustment using the $\mu$-defects at $\varepsilon=0,-1/(2\nu_i^2)$, where $\nu_i$ are the effective quantum numbers for the ab initio curves with $n=4,5,6,7,8$. 

While the $\eta$ defects avoid the issues related to unphysical interlopers, they show a stronger energy-dependence than the $\mu$ defects in the case of H$_2$. To mitigate this effect, the MQDT program uses functions of the form $A_f(\varepsilon)\eta_\text{f}^{(S\Lambda)}(R,\varepsilon)$, where $A_f(\varepsilon)$ is the Ham scaling factor, when constructing the MQDT matrices for a given energy.

\subsection{MQDT Fits of Spin-Rotation and Rotational Intervals}
\label{subsec:mqdt_fits}

\begin{figure}[h]\centering
	{\includegraphics[trim=0cm 0.0cm 0cm 0cm, clip=false, width=0.8\linewidth]{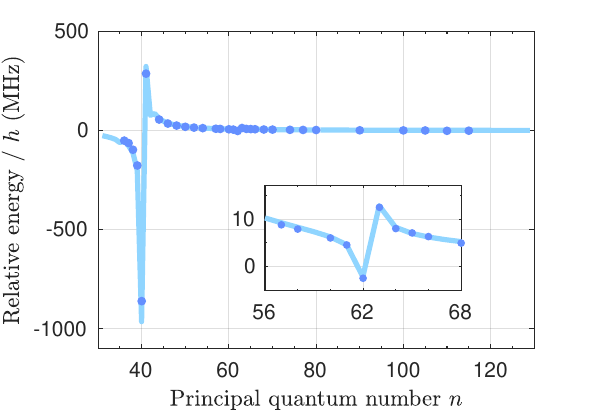}}
      \caption{MQDT-calculated (light blue line) and experimentally obtained (colored dots) binding energies of the middle fine-structure component (color coding corresponding to that shown in Fig. \ref{fig:levelstr}) of the $n$f2($J=1$) Rydberg states. All binding energies are referenced to the energies of the corresponding unperturbed $n$f series.}  
              \label{fig:7f_perturbation}	
\end{figure}

MQDT calculations based on the ab initio quantum-defect parameters introduced in Subsection~\ref{subsec:mqdt_theory} were used to determine the rotational and spin-rotational level structure in the X$^+(v^+=0)$ ground state of H$_2^+$ in a least-squares fit to the positions of the $n$f2($J=1$), $n$f4($J=1$) and $n$f6($J=3$) Rydberg states determined as explained in Section~\ref{sec:results}. As an example, Fig.~\ref{fig:7f_perturbation} displays the results of an MQDT calculation (light blue line) in comparison to the measured positions (blue dots) of the middle fine-structure component (color coding corresponding to that shown in Fig. \ref{fig:levelstr}) of the $n$f2($J=1$) states (see below for a discussion of the energy perturbations around $n$ = 40 and $n$ = 62). An iterative fitting procedure was adopted, with separate fits of the spin-rotational intervals of H$_2^+$ for fixed values of the pure (spin-free) rotational intervals, and of the spin-free rotational intervals for fixed values of the spin-rotational intervals. The advantages of this procedure for the determination of the H$_2^+$ spin-rotational intervals are that (i) the fine-structure splittings of the $n$f$N^+$ Rydberg states of H$_2$ are not sensitive to the exact positions of the H$_2^+$ rotational levels, as long as the series are free of perturbations from rotational channel interactions, (ii) the spin-rotational splitting in H$_2^+$ can be reliably derived from the fine-structure splittings of the $n$f Rydberg states without the need to know their absolute energies accurately, (iii) the transition frequencies needed to access different fine-structure components of a given $n$f$N^+$ state from a given intermediate state are almost identical and subject to the same first-order Doppler shifts, and finally (iv) the fine-structure components of a given $n$f$N^+$ state have similar polarizabilities and thus similar Stark shifts. As a result of (iii) and (iv), the residual Stark and first-order Doppler shifts and their associated uncertainties cancel out when determining the fine-structure splittings as differences between transitions frequencies to different fine-structure components of any given $n$f$N^+$ state. Reliable fine-structure intervals can therefore be determined at $n$ values beyond 100 and used to determine the H$_2^+$ spin-rotational intervals. The determination of the pure rotational energies of H$_2^+$ in an MQDT least-squares fit requires precise values of the absolute transition frequencies and does not benefit from these advantages, but is more efficient once the spin-rotational intervals of the H$_2^+$ rotational levels have been determined.

\begin{figure*}
	{\includegraphics[trim=0cm 0.0cm 0cm 0cm, clip=false, width=0.95\linewidth]{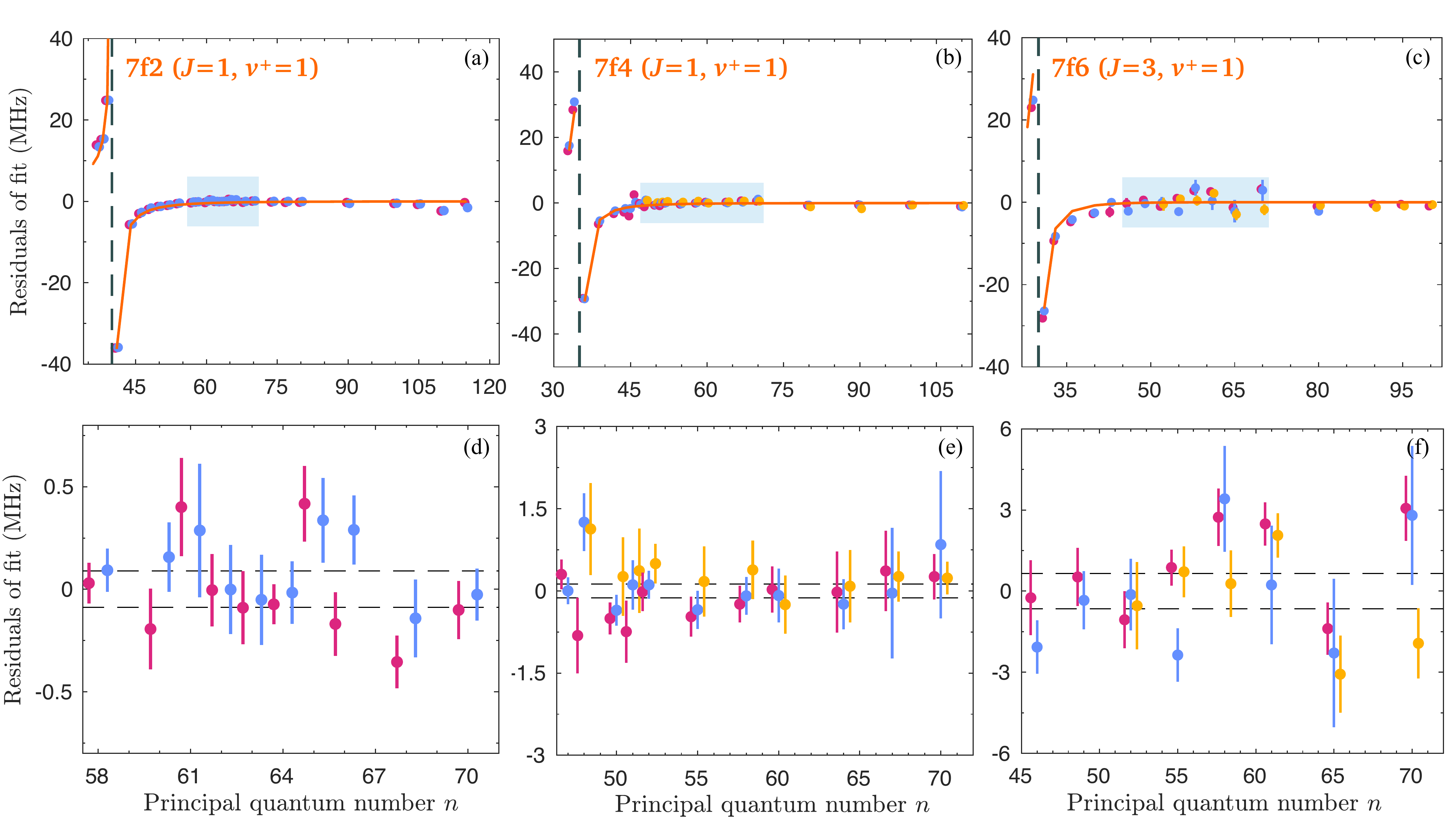}}
      \caption{Residuals of the binding energies X$^+$(0,2) $\leftarrow$ GK(0,0) [(a),(d)], X$^+$(0,4) $\leftarrow$ GK(0,2) [(b),(e)], and X$^+$(0,6) $\leftarrow$ I$^+$(0,4) [(c),(f)]. The color coding of the data points representing different fine-structure components corresponds to that shown in Fig. \ref{fig:levelstr}. Panels (a)-(c) show the entire range of $n$f states which were experimentally probed. The vertical grey lines indicate the positions of the 7f2$(J=1)$, 7f4$(J=1)$, and 7f6$(J=3)$ ($v^+=1$) Rydberg states, which cause perturbations in the residuals. The orange lines indicate the residuals obtained in a simple two-channel model of the interaction between the $n$f$N^+(v^+=0)$ series and the 7f$N^+(v^+=1)$ perturbers (see text for details). The blue rectangles in panels (a)-(c) show the selected ranges of principal quantum numbers which were used to extract the reported values of the X$^+$(0,2) $\leftarrow$ GK(0,0), X$^+$(0,4) $\leftarrow$ GK(0,2) and  X$^+$(0,6) $\leftarrow$ I$^+$(0,4) binding energies. The residuals obtained after correction with the two-channel model
      are shown on an expanded scale in panels (d)-(f), in which the horizontal dashed lines indicate the standard deviations of the datasets in each case.}  
              \label{fig:rotint_resi}	
\end{figure*}

The results of the least-squares fits are the extrapolated positions of the spin-rotational and rotational levels of H$_2^+$ with respect to the positions of the GK(0,0), GK(0,2) and I$^+$(0,4) intermediate states used to record the spectra of the $n$f2($J=1$), $n$f4($J=1$) and $n$f6($J=3$) Rydberg series. The energies of all fine-structure components of the measured $n$f$N^+$ Rydberg series with respect to their corresponding ionization limits are computed in an MQDT calculation including the effects of electron spins as explained in Ref.~\cite{haase15a}, see also \cite{jungen98a,jungen23a}. Adding these Rydberg-electron energies to each of the corresponding observed transition frequencies from the selected intermediate states yields independent values for the binding energy of that intermediate state. The residuals shown in Fig. \ref{fig:rotint_resi} are computed as the differences between these independent values and their weighted averages, which represent the X$^+$(0,2) $-$ GK(0,0), X$^+$(0,4) $-$ GK(0,2) and X$^+$(0,6) $-$ I$^+$(0,4) energy intervals. The weights of individual contributions are the inverse squares of the experimental uncertainties [see vertical error bars in Fig.~\ref{fig:rotint_resi}(d)-(f)].

Figs.~\ref{fig:rotint_resi}(a)-(c) present an overview of the residuals obtained from all measured transitions to $n$f2($J=1$), $n$f4($J=1$) and $n$f6($J=3$) Rydberg states, respectively. Most residuals are within the experimental uncertainties and significantly below 1~MHz, confirming the accuracy of the MQDT parameters expected from the considerations presented in Subsection~\ref{subsec:mqdt_theory}. However, residuals up to $\pm$ 40 MHz between measured and calculated positions are observed for each series near the calculated positions of the 7f($v^+=1$) Rydberg states of the corresponding $N^+$ and $J$ values, i.e., 7f2($J=1,v^+=1$), 7f4($J=1,v^+=1$) and 7f6($J=3,v^+=1$), as indicated by the dashed grey vertical lines. These low-$n$ states with vibrationally excited ion-cores cause strong perturbations of the Rydberg series through interactions between channels of the same $N^+$, $\ell$ and $J$ values but differing in $v^+$ by $\pm 1$. 

As explained in Subsection~\ref{subsec:mqdt_theory}, these channel interactions should be accurately described by the quantum-defect parameters derived from ab initio calculations. Indeed, Fig.~\ref{fig:7f_perturbation} shows that the MQDT calculations reproduce the $\approx$ 1 GHz experimental shifts caused by the 7f2($J=1,v^+=1$) perturber in the region around the 40f2($J=1,v^+=0$) state to a level of 95\% accuracy. The reason for the inability of the MQDT calculations to exactly reproduce the level structure close to the low-$n$ perturbing states does not lie in the inaccuracy of the quantum-defect parameters. Instead, it originates from a problem in the definition of the energy of the Rydberg electron arising from the energy dependence of the quantum defects when levels with ion cores in different rovibrational states interact \cite{greene85b,jungen11a}. In situations involving a single channel, or two channels associated with energetically close-lying ion-core states, the Rydberg electron has a well defined binding energy and the values of the quantum defects can be accurately determined using the energy-dependent quantum-defect parameters determined ab initio. As illustration, the inset of Fig.~\ref{fig:7f_perturbation} displays the $\approx$ 10 MHz energy shifts in the region around the 62f2($J=1$) state, which are caused by a rotational-channel interaction with the 16f4$(J=1)$ state. The MQDT calculations perfectly reproduce these shifts, as confirmed by Fig. \ref{fig:rotint_resi} (a), in which the residuals show no effect of the rotational perturber. In contrast, if two channels with very different ion-core energies interact [as is the case for the perturbations caused by the 7f ($v^+=1$) states], the Rydberg-electron binding energy, and thus the value of the energy-dependent quantum defect parameters, depend on the ion-core level used as origin of the energy scale \cite{greene85b,gao90a,hvizdos20a,jungen11a}. In standard MQDT calculations, this problem of the energy scale is avoided by referencing the electron binding energy to the same energy (the mean of the two ion-core energies) when determining the values of the quantum defect parameters. This procedure, which makes the MQDT formalism efficient, is an approximation. It is physically motivated, in electron-(positive) ion systems, by the strongly attractive Coulomb potential which results in the electron having similar kinetic energies in the core region for channels associated with different ion-core states \cite{jungen11a}. The effects of this approximation are usually not detected, either because the quantum-defect parameters and their energy dependence are not known with sufficient precision in the first place, or because the Rydberg spectra are not sufficiently resolved. In the present study, both the quantum-defect parameters and the experimental data in the energy regions close to the $v^+=1$ perturbers are of high accuracy, thus providing an ideal opportunity for a detailed characterization of the effects of the energy-reference ambiguity. The data shown in Fig.\ref{fig:7f_perturbation} could therefore be used to test the performance of alternative vibrational frame transformations as introduced in Ref.~\cite{greene85b,gao90a,hvizdos20a}.

Inspection of the functional form of the energy dependence of the residuals in Fig.~\ref{fig:rotint_resi}(a)-(c) leads to the conclusions that (1) the effect of the vibrational-channel interaction is the same for all fine-structure components of a given state of principal quantum number $n$, so that the spin fine structure can be neglected in the following discussion;
and (2) they closely reproduce the functional form of the level shifts caused by the perturbation (see Fig.~\ref{fig:7f_perturbation}).
Consequently, the residuals can be accurately modeled using a simple two-channel interaction model, one consisting of an isolated resonance at the position of the low-$n$ perturber, and the other of a Rydberg series with positions corresponding to the unperturbed positions of the high-$n$ levels of the f series. Such a model involves three adjustable parameters: two quantum defects, one ($\eta_{11}$) to describe the position of the perturbing 7f$N^+(v^+=1)$ level and the other ($\eta_{00}$) to describe the positions of the $n$f$N^+(v^+=0)$ series, as well as one channel-interaction parameter, an off-diagonal quantum defect $\eta_{10}$, which determines the overall amplitude of the residuals. The orange lines in Fig.~\ref{fig:rotint_resi}(a)-(c) depict the residuals calculated with this two-channel model after optimization of the three parameters to the values of $\eta_{00}$ = 1.0 $\times$ 10$^{-5}$, $\eta_{11}$ = 2.3 $\times$ 10$^{-2}$, and $\eta_{10}$ = 1.41 $\times$ 10$^{-3}$. They perfectly describe the residuals of the MQDT calculations (colored dots). Moreover, the three sets of residuals can be modelled with the same set of three parameters, as expected for channel interactions that are all $\Delta v^+=1$, $\Delta N^+=0$, $\Delta \ell=0$ interactions involving a 7f($v^+=1$) perturbing level and high $n$f($v^+=0$) Rydberg states. Fig.~\ref{fig:rotint_resi} thus demonstrates that the energy-reference inconsistency of standard MQDT calculations can be corrected empirically using a two-channel interaction model. 

Figs.~\ref{fig:rotint_resi}(d)-(f) present the residuals after this correction on an expanded scale in the range of principal quantum numbers used to determine the rotational and spin-rotational level structure of H$_2^+$. States with principal quantum number beyond 70 were excluded from this analysis to limit the systematic uncertainties from dc Stark shifts caused by residual stray electric fields to less than 150 kHz, i.e., well within the statistical uncertainty of the measurements. 
The residuals do not show any trends over the displayed ranges of principal quantum numbers $n$ and their distribution does not strongly deviate from a normal distribution. We therefore conclude that the combination of MQDT calculations based on the new set of ab initio quantum-defect parameters with an empirical two-channel interaction model to correct for the energy-reference inconsistency of standard MQDT near perturbations involving low-$n$ Rydberg states can be used to determine ionization energies by extrapolation at a precision limited by the experimental uncertainties in the transition frequencies. 

The upper part of Table \ref{tab:rotint} summarizes all energy intervals relevant to the determination of the spin-free term values of the rotational levels of the ground vibronic state of para-H$_2^+$ and their uncertainties. The lower part lists the experimental values of the three lowest rotational intervals of para-H$_2^+$, which are determined here for the first time with such precision. More precise ab initio results by Korobov \cite{korobov06a,korobov08a} are also available for the two lowest rotational intervals. The experimental values obtained in this work are in excellent agreement, at the level of 1$\sigma$, with these ab initio results.

\begin{table*}
\centering
\caption{Summary of energy intervals used to determine the first three rotational intervals in para-H$_{2}^+$ and their uncertainties.}
\label{tab:rotint}
\begin{tabular}{llll} 
\toprule
& \ \ \ Energy interval & \ \ \ Value (cm$^{-1}$) &\ \ \ Ref. \\
\midrule
(1) & \ \ \ X$^+ (0,0)$ $-$ GK(0,2)  \  & \ \ \ 12\,723.757\,440\,7(18)$_\textrm{stat}$(40)$_\textrm{sys}$  & \ \ \ \cite{doran24a} \\
(2) & \ \ \ X$^+ (0,2)$ $-$ GK(0,0)  \ & \ \ \ 
12\,962.876\,609\,0(30)$_\textrm{stat}$ & \ \ \ This work \\
(3) & \ \ \ X$^+ (0,4)$ $-$ GK(0,2)  & \ \ \  13\,299.213\,073\,2(42)$_\textrm{stat}$  & \ \ \ This work \\
(4) & \ \ \ X$^+ (0,6)$ $-$ I$^+$(0,4)  & \ \ \ 12\,905.184\,142(22)$_\textrm{stat}$(170)$_\textrm{sys}$ & \ \ \ This work \\
(5) & \ \ \ GK(0,2) $-$ GK(0,0) & \ \ \ \ \ \ \ \, 64.882\,423\,7(11)$_\textrm{stat}$ & \ \ \ \cite{hoelsch18a} \\
(6) & \ \ \ I$^+$(0,4) $-$ GK(0,2) &  \ \ \ \ \, 1009.958\,87(7)$_\textrm{stat}$ & \ \ \ This work \\
(7) = (5)+(6) & \ \ \ I$^+$(0,4) $-$ GK(0,0) & \ \ \ \ \ \ \, 1074.841\,29(7)$_\textrm{stat}$  & \ \ \ This work \\
(8) = (2)$-$(1)$-$(5) & \ \ \ X$^+$(0,2) $-$ X$^+$(0,0) & \ \ \ \ \ \ \, 174.236\,744\,6(77) & \ \ \ This work  \\
(9)& \ \ \ X$^+$(0,2) $-$ X$^+$(0,0) & \ \ \ \ \ \ \, 174.236\,742\,482\,36 & \ \ \ Theory \cite{korobov06a, korobov08a} \\
(10)& \ \ \ X$^+$(0,2) $-$ X$^+$(0,0) & \ \ \ \ \ \ \, 174.236\,71(7) & \ \ \ \cite{haase15a} \\
(11) = (3)$-$(1) & \ \ \ X$^+$(0,4) $-$ X$^+$(0,0) & \ \ \ \ \ \ \, 575.455\,632\,5(86) & \ \ \ This work  \\
(12)& \ \ \ X$^+$(0,4) $-$ X$^+$(0,0) & \ \ \ \ \ \ \, 575.455\,640\,795\,85 & \ \ \ Theory \cite{korobov06a, korobov08a} \\
(13) = (4)$-$(1)$+$(6) & \ \ \ X$^+$(0,6) $-$ X$^+$(0,0) & \ \ \ \ \, 1191.385\,571(240) & \ \ \ This work \\

\bottomrule  
\end{tabular} 
\end{table*}

\begin{figure*}
	{\includegraphics[trim=0cm 0.0cm 0cm 0cm, clip=false, width=0.95\linewidth]{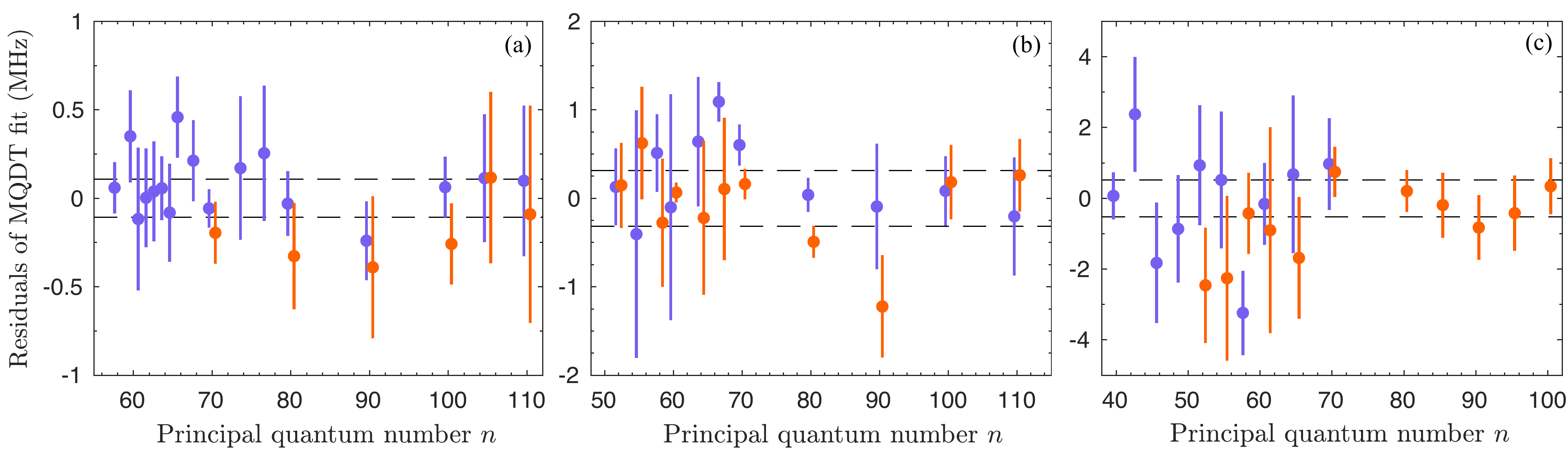}}
      \caption{Residuals of the fitted energy intervals $\Delta_1$ (purple dots) and $\Delta_2$ (orange dots) (see also Fig. \ref{fig:levelstr}), displayed as the difference between the experimentally determined and MQDT-predicted energy intervals, for $n$f Rydberg states converging to the $N^+=2$ (a), $N^+=4$ (b), and $N^+=6$ (c) ionization thresholds. The horizontal dashed lines indicate the standard deviations of the data sets.}  
              \label{fig:spinrot_resi}	
\end{figure*}

Many more transitions, to states with $n$ values up to 115 (see, e.g., Fig.~\ref{fig:Np2_n110_spectrum}), could be used to determine the H$_2^+$ spin-rotation intervals 
$\Delta E_{\textrm{SR}}(N^+=2)$, $\Delta E_{\textrm{SR}}(N^+=4)$ and $\Delta E_{\textrm{SR}}(N^+=6)$, as explained above. In the fits, the least-squares deviations between the experimentally determined splittings $\Delta_1$ and $\Delta_2$ (see Fig. \ref{fig:levelstr}) between fine-structure components and the corresponding energy splittings calculated by MQDT for varying values of the spin-rotation coupling constant were minimized. Fig. \ref{fig:spinrot_resi} displays the fit residuals corresponding to $\Delta_1$ in purple and to $\Delta_2$ in orange, with the dashed lines indicating the standard deviations of the data sets. The increased error bars and the larger scatter of the experimental data obtained for the spin-rotation splitting of the $N^+=6$ level are the result of the poor signal-to-noise ratio of these spectra. 
\par
The spin-rotation coupling constants $\gamma_{\textrm{SR}}^{N^+=2}$, $\gamma_{\textrm{SR}}^{N^+=4}$ and $\gamma_{\textrm{SR}}^{N^+=6}$ are calculated from the optimized spin-rotation intervals using Eq.~\eqref{eq:spinrot_int}. $\gamma_{\textrm{SR}}^{(0)}$ and $\gamma_{\textrm{SR}}^{(1)}$ are determined in a least-squares fit from $\gamma_{\textrm{SR}}^{N^+=2}$, $\gamma_{\textrm{SR}}^{N^+=4}$ and $\gamma_{\textrm{SR}}^{N^+=6}$ using Eq. \ref{eq:gamma_dep}. The residuals of this fit are below 100 kHz, i.e. on the order of the uncertainties of $\gamma_{\textrm{SR}}^{N^+=2}$, $\gamma_{\textrm{SR}}^{N^+=4}$ and $\gamma_{\textrm{SR}}^{N^+=6}$, confirming the validity of Eq.~\eqref{eq:gamma_dep}. Table \ref{tab:spinrot} lists all obtained values, which are determined here for the first time with a precision below 100 kHz for each of the $N^+=2$, 4 and 6 rotational levels. They are in agreement, within their uncertainties, with the more precise ab initio theoretical determinations \cite{korobov06c,haidar22a}. 
\begin{table}[h]
\centering
\caption{Summary of spin-rotation coupling constants $\gamma_{\textrm{SR}}$ for the three lowest rotational levels of para-H$_2^+$ and  $\gamma_{\textrm{SR}}^{(0)}$ and $\gamma_{\textrm{SR}}^{(1)}$ obtained from Eq. \ref{eq:gamma_dep}.}
\label{tab:spinrot}
\begin{tabular}{llll} 
\toprule
$\gamma_{\textrm{SR}}$ & \ \ \ Value (MHz) & \ \ \ Ref. \\
\midrule
$\gamma_{\textrm{SR}}^{N^+=2}$ & \ \ \ 42.21(4)  & \ \ \ This work \\
$\gamma_{\textrm{SR}}^{N^+=2}$ & \ \ \ 42.16352(15) & \ \ \ Theory \cite{haidar22a} \\
$\gamma_{\textrm{SR}}^{N^+=4}$  & \ \ \ 41.26(8)  & \ \ \ This work \\
$\gamma_{\textrm{SR}}^{N^+=4}$  & \ \ \ 41.2942  & \ \ \ Theory \cite{korobov06c} \\
$\gamma_{\textrm{SR}}^{N^+=6}$ &  \ \ \ 40.04(8)  & \ \ \ This work\\
$\gamma_{\textrm{SR}}^{(0)}$ & \ \ \ 42.53(9) & \ \ \ This work \\
$\gamma_{\textrm{SR}}^{(1)}$ & \ \ $-$0.060(3) & \ \ \ This work \\

\bottomrule
\end{tabular} 
\end{table}

\section{Conclusions}
In this article, we have presented a study of the $n$f Rydberg series of H$_2$ converging on the low-lying $N^+=2$, 4 and 6 rotational levels of of the X$^+$ $^2\Sigma_g^+(v^+=0)$ vibronic ground state of para-H$_2^+$. Using single-mode and frequency-comb-calibrated continuous-wave laser radiation, we have recorded spectra of Rydberg states over a broad range of principal quantum number $n$, from below 40 to 115. These spectra reveal, with full details, how the fine structure of the Rydberg states gradually evolves from a situation where the interactions determining the level patterns are the short-range electrostatic exchange interaction and the long-range electrostatic interactions between the Rydberg electron and the ion core, to a situation where the spin-rotation interaction in the ionic core determines the main fine-structure splittings.

In the former situation, the Rydberg states form two groups of levels of well-defined total electron spin ($S=0$ or 1). Each group displays splittings characteristic of the long-range electrostatic interactions that couple the Rydberg-electron orbital angular momentum $\vec{\ell}$ to the H$_2^+$-ion-core rotational angular momentum $\vec{N^+}$ to form a nearly conserved total angular momentum $\vec{N}$. In the latter situation, the Rydberg electron is decoupled from the ion core so that the main splitting of the fine-structure pattern corresponds to the spin-rotation interval of the H$_2^+$ ion core.

The observed gradual uncoupling of the Rydberg electron with increasing $n$ value was quantitatively analyzed in MQDT calculations using quantum defects derived from recent high-level ab initio calculations of low-$n$ Rydberg states of H$_2$. These calculations were used to extrapolate the Rydberg series to their limits and to determine, for the first time, the rotational and spin-rotational structures of the ground vibronic state of para-H$_2^+$ with sub-MHz accuracy.

Comparing the experimentally measured positions of several $n$f series with the positions calculated by MQDT revealed an overall excellent agreement, within the experimental uncertainties, but also significant discrepancies in the vicinity of low-$n$ perturbing Rydberg states with a vibrationally excited H$_2^+$ ion core. These discrepancies could be attributed to the energy-reference ambiguity of standard MQDT and quantitatively accounted for using a simple empirical model involving two interacting channels. In future, a detailed analysis of these discrepancies may serve as benchmarks for extensions of MQDT beyond the approximation of channel-independent reference electron energies.

The combination of precision Rydberg spectroscopy with MQDT is demonstrated to be a very attractive way to achieve a simultaneously global and precise description of molecular photoionization and to accurately determine the energies of a broad range of electronic, vibrational, and rotational levels of molecular ions, including their fine and hyperfine structures.

\begin{acknowledgments}
We thank Dr. Christian Jungen, Orsay, for his invaluable assistance and discussions concerning all aspects of multichannel quantum-defect theory and also for allowing us to use his programs. We also thank Hansj\"urg Schmutz and Josef A. Agner for their technical assistance. M.B. acknowledges NWO for a VENI grant (VI.Veni.202.140). This work is supported financially by the Swiss National Science Foundation under the excellence grant No. 200020B-200478.
\end{acknowledgments}

\newpage

%

%

\end{document}


\renewcommand{\thetable}{S\arabic{table}}
\renewcommand{\thefigure}{S\arabic{figure}}
\newpage

\KOMAoptions{paper=portrait,pagesize}
\recalctypearea

\section{Supplemental Material}

\begin{figure}[h]\centering
	{\includegraphics[trim=0cm 0.0cm 0cm 0cm, clip=false, width=0.98\linewidth]{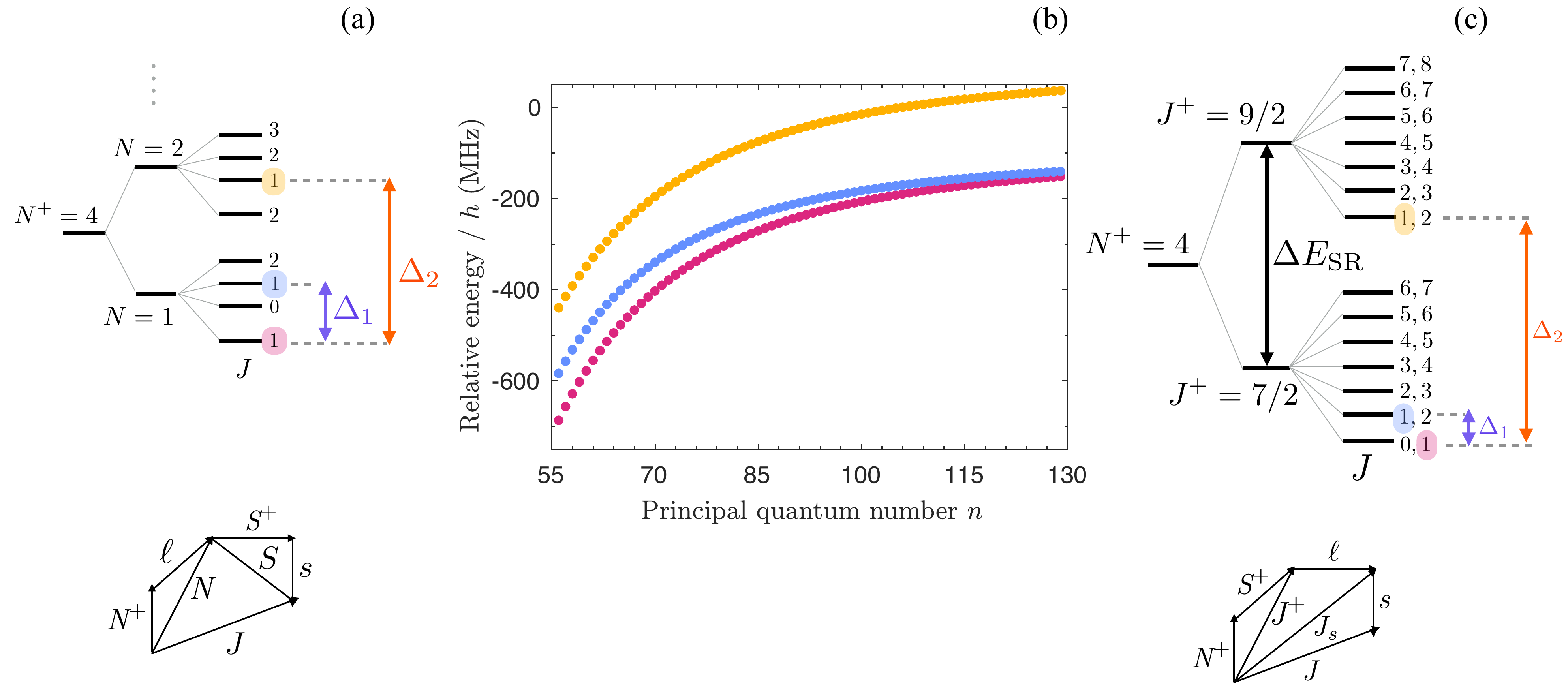}}
      \caption{Positions of the $n$f4$(J=1)$ Rydberg states of H$_2$, obtained in an MQDT calculation including electron spins, relative to the zero-quantum-defect positions $-cR_{\textrm{H}_2}$/$n^2$ (panel (b)).}  
              \label{fig:levelstrNp4}	
\end{figure}

\begin{figure}[h]\centering
	{\includegraphics[trim=0cm 0.0cm 0cm 0cm, clip=false, width=0.98\linewidth]{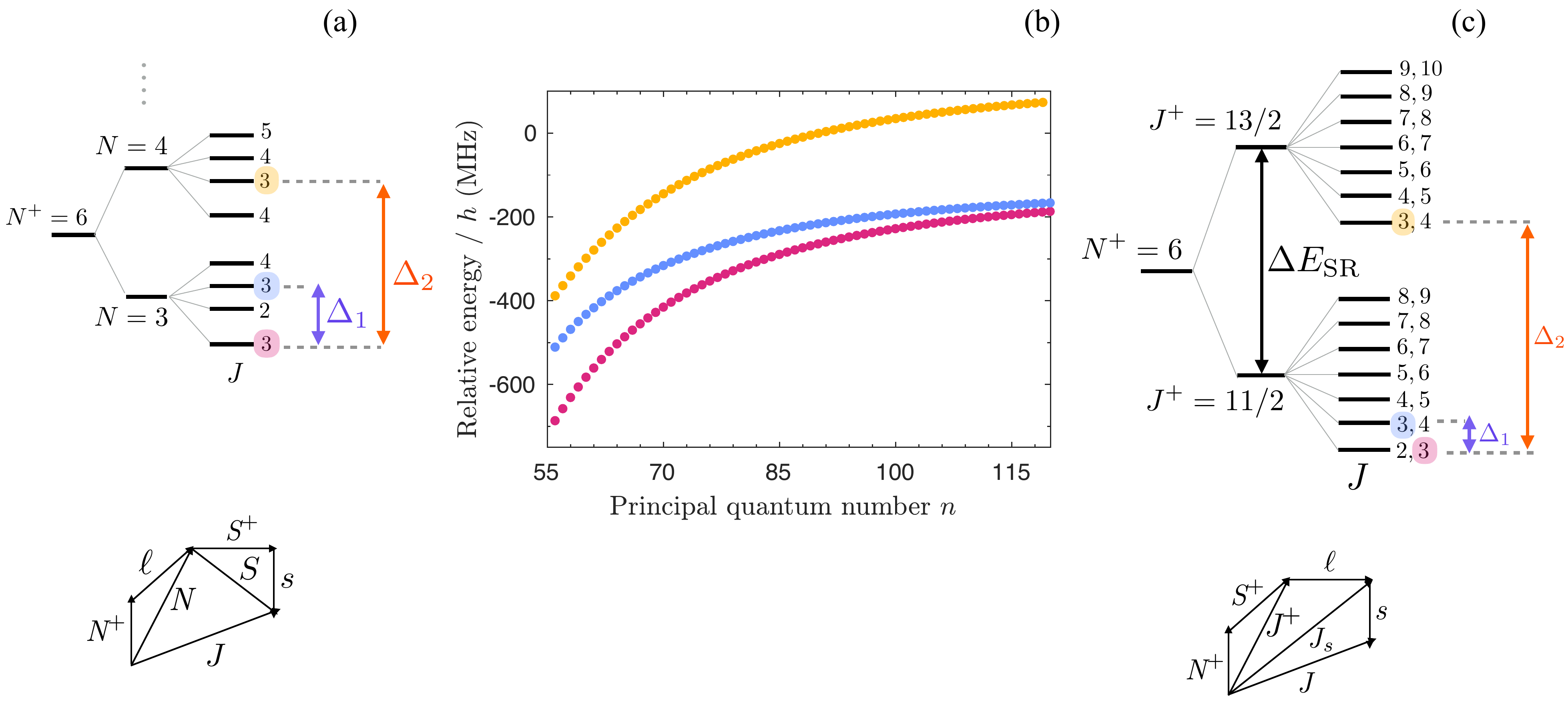}}
      \caption{Positions of the $n$f6$(J=3)$ Rydberg states of H$_2$, obtained in an MQDT calculation including electron spins, relative to the zero-quantum-defect positions $-cR_{\textrm{H}_2}$/$n^2$ (panel (b)). }  
              \label{fig:levelstrNp6}	
\end{figure}

\pagenumbering{arabic}
This supplemental material provides information on \\

a) the energy-level structure of $n$f4($J=1$) and $n$f6($J=3$) Rydberg states of H$_2$ determined in MQDT calculations including electron spins, as described for $n$f2($J$=1) Rydberg states in the article (Figs. \ref{fig:levelstrNp4} and \ref{fig:levelstrNp6}); \\ 

b) the measured positions of the $n$f Rydberg states of H$_2$ and their binding energies determined experimentally and by MQDT (before and after correcting with an empirical two-channel interaction model) using the fitted values of the spin-rotation coupling constants and of the X$^+ (0,2) - $GK$(0,0) $, X$^+ (0,4) - $GK$(0,2) $ and X$^+ (0,6) - $I$^+(0,4) $ intervals (Tables \ref{tab:meas_fr_Np2}, \ref{tab:meas_fr_Np4}, \ref{tab:meas_fr_Np6}).

\newpage

\KOMAoptions{paper=portrait,pagesize}
\recalctypearea

\newpage

\KOMAoptions{paper=landscape,pagesize}{}

\begin{longtable}{lrrrrrrr}

\caption{Experimental positions (in wavenumber units cm$^{-1}$) of the measured fine-structure components of the $n$f2$(J=1)$ Rydberg states of H$_2$ relative to the GK(0,0) state, their statistical uncertainties (in MHz), their binding energies determined using the measured positions and the fitted $\gamma_{\textrm{SR}}^{N^+=2}$ and X$^+ (0,2) - $GK$(0,0)$ interval, their binding energies calculated by MQDT, their binding energies calculated by MQDT after correcting with a two-channel interaction model, the deviation between experiment and MQDT calculations, and the deviation between experiment and MQDT after correcting with a two-channel interaction model. The states are labelled by the principal quantum number $n$ and by the letter A, B or C to denote the different fine-structure components of $J$=1 in the order of increasing energy at each value of $n$.} \\

\toprule
$n$(Label) & \ \  $\tilde\nu_{\textrm{obs}}$ (cm$^{-1}$) & \ \  \ \ $\sigma$ (MHz) & \ \  $E_{\textrm{bind}} ^{\textrm{obs}} \ / \ hc$ (cm$^{-1}$) & \ \  $E_{\textrm{bind}} ^{\textrm{MQDT}} \ / \ hc$ (cm$^{-1}$) 
& \ \  $E_{\textrm{bind}} ^{\textrm{corr}} \ / \ hc$ (cm$^{-1}$) 
& \ \  $\nu_{\textrm{obs}} - \nu_{\textrm{MQDT}}$ (MHz) 
& \ \   $\nu_{\textrm{obs}} - \nu_{\textrm{corr}}$ (MHz)  \\
\midrule
36(A)& \ \ 12\,878.152\,504\,4& \ \  0.523& \ \ $-$84.724\,104\,5& \ \ $-$84.723\,857\,4& \ \ $-$84.723\,538\,4& \ \ $-$7.409& \ \ $-$16.972 \\
36(B)& \ \ 12\,878.154\,913\,1& \ \ 0.616& \ \ $-$84.721\,695\,8& \ \ $-$84.721\,747\,8& \ \ $-$84.721\,428\,8& \ \ \, \ 1.557& \ \ $-$8.004 \\
37(A)& \ \ 12\,882.671\,851\,9& \ \ 0.327& \ \ $-$80.204\,757\,0& \ \ $-$80.205\,218\,8& \ \ $-$80.204\,837\,2& \ \ 13.843& \ \ 2.404 \\
37(B)& \ \ 12\,882.673\,956\,1& \ \ 0.269& \ \ $-$80.202\,652\,8& \ \ $-$80.203\,099\,2& \ \ $-$80.202\,717\,6& \ \ 13.381& \ \ 1.943 \\
38(A)& \ \ 12\,886.837\,940\,7& \ \ 0.429& \ \ $-$76.038\,668\,2& \ \ $-$76.039\,174\,5& \ \ $-$76.038\,670\,4& \ \ 15.177& \ \ 0.063 \\
38(B)& \ \ 12\,886.840\,072\,8& \ \ 0.448& \ \ $-$76.036\,536\,1& \ \ $-$76.037\,048\,6& \ \ $-$76.036\,544\,5& \ \ 15.363& \ \ 0.250 \\
39(A)& \ \ 12\,890.685\,989\,5& \ \ 0.511& \ \ $-$72.190\,619\,5& \ \ $-$72.191\,446\,8& \ \ $-$72.190\,619\,0& \ \ 24.803& \ \ $-$0.014 \\
39(B)& \ \ 12\,890.688\,110\,6& \ \ 0.381& \ \ $-$72.188\,498\,4& \ \ $-$72.189\,326\,9& \ \ $-$72.188\,499\,1& \ \ 24.839& \ \ 0.024 \\
40(A)& \ \ 12\,894.227\,923\,9& \ \ 0.261& \ \ $-$68.648\,685\,1& \ \ $-$68.651\,506\,9& \ \ $-$68.647\,892\,3& \ \ 84.596& \ \ $-$23.765 \\
40(B)& \ \ 12\,894.230\,707\,0& \ \ 0.462& \ \ $-$68.645\,902\,0& \ \ $-$68.649\,359\,1& \ \ $-$68.645\,744\,7& \ \ 103.642& \ \ $-$4.714 \\
41(A)& \ \ 12\,897.574\,332\,0& \ \ 0.367& \ \ $-$65.302\,277\,0& \ \ $-$65.301\,074\,7& \ \ $-$65.302\,272\,6& \ \ $-$36.043& \ \ $-$0.132 \\
41(B)& \ \ 12\,897.576\,588\,4& \ \ 0.287& \ \ $-$65.300\,020\,6& \ \ $-$65.298\,823\,3& \ \ $-$65.300\,021\,1& \ \ $-$35.893& \ \ 0.016 \\
44(A)& \ \ 12\,906.171\,907\,6& \ \ 0.164& \ \ $-$56.704\,701\,4& \ \ $-$56.704\,510\,2& \ \ $-$56.704\,691\,8& \ \ $-$5.731& \ \ $-$0.286 \\
44(B)& \ \ 12\,906.174\,130\,8& \ \ 0.159& \ \ $-$56.702\,478\,2& \ \ $-$56.702\,293\,3& \ \ $-$56.702\,474\,9& \ \ $-$5.542& \ \ $-$0.097 \\
\newpage
46(A)& \ \ 12\,910.996\,478\,3& \ \ 0.107& \ \ $-$51.880\,130\,7& \ \ $-$51.880\,029\,8& \ \ $-$51.880\,128\,6& \ \ $-$3.024& \ \ $-$0.062 \\
46(B)& \ \ 12\,910.998\,730\,1& \ \ 0.090& \ \ $-$51.877\,878\,9& \ \ $-$51.877\,788\,3& \ \ $-$51.877\,887\,1& \ \ $-$2.716& \ \ 0.246 \\
48(A)& \ \ 12\,915.230\,709\,4& \ \ 0.075& \ \ $-$47.645\,899\,6& \ \ $-$47.645\,833\,8& \ \ $-$47.645\,894\,4& \ \ $-$1.972& \ \ $-$0.154 \\
48(B)& \ \ 12\,915.232\,987\,2& \ \ 0.094& \ \ $-$47.643\,621\,8& \ \ $-$47.643\,567\,3& \ \ $-$47.643\,627\,9& \ \ $-$1.633& \ \ 0.185 \\
50(A)& \ \ 12\,918.967\,018\,8& \ \ 0.076& \ \ $-$43.909\,590\,2& \ \ $-$43.909\,547\,8& \ \ $-$43.909\,587\,1& \ \ $-$1.271& \ \ $-$0.091 \\
50(B)& \ \ 12\,918.969\,312\,1& \ \ 0.095& \ \ $-$43.907\,296\,9& \ \ $-$43.907\,255\,4& \ \ $-$43.907\,294\,7& \ \ $-$1.244& \ \ $-$0.064 \\
52(A)& \ \ 12\,922.280\,477\,9& \ \ 0.091& \ \ $-$40.596\,131\,1& \ \ $-$40.596\,098\,1& \ \ $-$40.596\,124\,2& \ \ $-$0.989& \ \ $-$0.206 \\
52(B)& \ \ 12\,922.282\,801\,3& \ \ 0.087& \ \ $-$40.593\,807\,7& \ \ $-$40.593\,778\,7& \ \ $-$40.593\,804\,8& \ \ $-$0.869& \ \ $-$0.087 \\
54(A)& \ \ 12\,925.232\,566\,4& \ \ 0.093& \ \ $-$37.644\,042\,6& \ \ $-$37.644\,023\,5& \ \ $-$37.644\,040\,8& \ \ $-$0.572& \ \ $-$0.055 \\
54(B)& \ \ 12\,925.234\,918\,7& \ \ 0.101& \ \ $-$37.641\,690\,3& \ \ $-$37.641\,675\,9& \ \ $-$37.641\,693\,2& \ \ $-$0.431& \ \ 0.086 \\
57(A)& \ \ 12\,929.091\,601\,3& \ \ 0.079& \ \ $-$33.785\,007\,7& \ \ $-$33.784\,995\,7& \ \ $-$33.785\,004\,4& \ \ $-$0.360& \ \ $-$0.100 \\
57(B)& \ \ 12\,929.094\,000\,6& \ \ 0.092& \ \ $-$33.782\,608\,4& \ \ $-$33.782\,603\,1& \ \ $-$33.782\,611\,8& \ \ $-$0.159& \ \ 0.101 \\
58(A)& \ \ 12\,930.246\,789\,4& \ \ 0.100& \ \ $-$32.629\,819\,6& \ \ $-$32.629\,814\,0& \ \ $-$32.629\,820\,6& \ \ $-$0.168& \ \ 0.030 \\
58(B)& \ \ 12\,930.249\,200\,1& \ \ 0.105& \ \ $-$32.627\,408\,9& \ \ $-$32.627\,405\,4& \ \ $-$32.627\,412\,0& \ \ $-$0.105& \ \ 0.093 \\
60(A)& \ \ 12\,932.386\,225\,2& \ \ 0.198& \ \ $-$30.490\,383\,8& \ \ $-$30.490\,374\,1& \ \ $-$30.490\,377\,3& \ \ $-$0.291& \ \ $-$0.194 \\
60(B)& \ \ 12\,932.388\,681\,0& \ \ 0.170& \ \ $-$30.487\,928\,0& \ \ $-$30.487\,930\,0& \ \ $-$30.487\,933\,2& \ \ 0.060& \ \ 0.157 \\
61(A)& \ \ 12\,933.377\,873\,3& \ \ 0.240& \ \ $-$29.498\,735\,7& \ \ $-$29.498\,747\,2& \ \ $-$29.498\,749\,1& \ \ 0.345& \ \ 0.401 \\
61(B)& \ \ 12\,933.380\,336\,4& \ \ 0.326& \ \ $-$29.496\,272\,6& \ \ $-$29.496\,280\,3& \ \ $-$29.496\,282\,2& \ \ 0.231& \ \ 0.287 \\
62(A)& \ \ 12\,934.321\,690\,5& \ \ 0.177& \ \ $-$28.554\,918\,5& \ \ $-$28.554\,917\,7& \ \ $-$28.554\,918\,4& \ \ $-$0.024& \ \ $-$0.004 \\
62(B)& \ \ 12\,934.324\,198\,4& \ \ 0.217& \ \ $-$28.552\,410\,6& \ \ $-$28.552\,409\,9& \ \ $-$28.552\,410\,6& \ \ $-$0.021& \ \ $-$0.001 \\
63(A)& \ \ 12\,935.221\,799\,4& \ \ 0.178& \ \ $-$27.654\,809\,6& \ \ $-$27.654\,807\,0& \ \ $-$27.654\,806\,6& \ \ $-$0.078& \ \ $-$0.090 \\
\newpage
63(B)& \ \ 12\,935.224\,182\,8& \ \ 0.220& \ \ $-$27.652\,426\,2& \ \ $-$27.652\,424\,9& \ \ $-$27.652\,424\,5& \ \ $-$0.039& \ \ $-$0.051 \\
64(A)& \ \ 12\,936.079\,208\,0& \ \ 0.097& \ \ $-$26.797\,401\,0& \ \ $-$26.797\,399\,9& \ \ $-$26.797\,398\,5& \ \ $-$0.033& \ \ $-$0.073 \\
64(B)& \ \ 12\,936.081\,676\,6& \ \ 0.153& \ \ $-$26.794\,932\,4& \ \ $-$26.794\,933\,2& \ \ $-$26.794\,931\,8& \ \ 0.024& \ \ $-$0.016 \\
65(A)& \ \ 12\,936.897\,511\,0& \ \ 0.185& \ \ $-$25.979\,098\,0& \ \ $-$25.979\,114\,1& \ \ $-$25.979\,111\,9& \ \ 0.483& \ \ 0.417 \\
65(B)& \ \ 12\,936.900\,001\,9& \ \ 0.206& \ \ $-$25.976\,607\,1& \ \ $-$25.976\,620\,5& \ \ $-$25.976\,618\,3& \ \ 0.402& \ \ 0.336 \\
66(A)& \ \ 12\,937.678\,879\,1& \ \ 0.155& \ \ $-$25.197\,729\,9& \ \ $-$25.197\,727\,2& \ \ $-$25.197\,724\,3& \ \ $-$0.081& \ \ $-$0.169 \\
66(B)& \ \ 12\,937.681\,408\,3& \ \ 0.169& \ \ $-$25.195\,200\,7& \ \ $-$25.195\,213\,3& \ \ $-$25.195\,210\,4& \ \ 0.378& \ \ 0.290 \\
68(A)& \ \ 12\,939.139\,492\,5& \ \ 0.129& \ \ $-$23.737\,116\,5& \ \ $-$23.737\,108\,9& \ \ $-$23.737\,104\,7& \ \ $-$0.228& \ \ $-$0.354 \\
68(B)& \ \ 12\,939.142\,049\,3& \ \ 0.190& \ \ $-$23.734\,559\,7& \ \ $-$23.734\,559\,2& \ \ $-$23.734\,555\,0& \ \ $-$0.015& \ \ $-$0.142 \\
70(A)& \ \ 12\,940.476\,701\,8& \ \ 0.142& \ \ $-$22.399\,907\,2& \ \ $-$22.399\,909\,1& \ \ $-$22.399\,903\,8& \ \ 0.057& \ \ $-$0.101 \\
70(B)& \ \ 12\,940.479\,287\,8& \ \ 0.127& \ \ $-$22.397\,321\,2& \ \ $-$22.397\,325\,6& \ \ $-$22.397\,320\,3& \ \ 0.132& \ \ $-$0.026 \\
74(A)& \ \ 12\,942.833\,114\,7& \ \ 0.236& \ \ $-$20.043\,494\,3& \ \ $-$20.043\,488\,7& \ \ $-$20.043\,481\,9& \ \ $-$0.168& \ \ $-$0.373 \\
74(B)& \ \ 12\,942.835\,769\,7& \ \ 0.330& \ \ $-$20.040\,839\,3& \ \ $-$20.040\,839\,5& \ \ $-$20.040\,832\,7& \ \ 0.006& \ \ $-$0.199 \\
77(A)& \ \ 12\,944.364\,657\,1& \ \ 0.228& \ \ $-$18.511\,951\,9& \ \ $-$18.511\,943\,5& \ \ $-$18.511\,935\,8& \ \ $-$0.252& \ \ $-$0.483 \\
77(B)& \ \ 12\,944.367\,362\,9& \ \ 0.308& \ \ $-$18.509\,246\,1& \ \ $-$18.509\,246\,2& \ \ $-$18.509\,238\,5& \ \ 0.003& \ \ $-$0.228 \\
80(A)& \ \ 12\,945.727\,123\,3& \ \ 0.259& \ \ $-$17.149\,485\,7& \ \ $-$17.149\,478\,7& \ \ $-$17.149\,470\,3& \ \ $-$0.210& \ \ $-$0.461 \\
80(B)& \ \ 12\,945.729\,873\,4& \ \ 0.182& \ \ $-$17.146\,735\,6& \ \ $-$17.146\,734\,4& \ \ $-$17.146\,726\,0& \ \ $-$0.036& \ \ $-$0.287 \\
90(A)& \ \ 12\,949.326\,519\,0& \ \ 0.138& \ \ $-$13.550\,090\,0& \ \ $-$13.550\,080\,6& \ \ $-$13.550\,070\,8& \ \ $-$0.282& \ \ $-$0.575 \\
90(B)& \ \ 12\,949.329\,399\,9& \ \ 0.175& \ \ $-$13.547\,209\,1& \ \ $-$13.547\,191\,8& \ \ $-$13.547\,182\,0& \ \ $-$0.519& \ \ $-$0.812 \\
100(A)& \ \ 12\,951.900\,979\,7& \ \ 0.123& \ \ $-$10.975\,629\,3& \ \ $-$10.975\,611\,9& \ \ $-$10.975\,601\,4& \ \ $-$0.522& \ \ $-$0.837 \\
100(B)& \ \ 12\,951.903\,992\,9& \ \ 0.120& \ \ $-$10.972\,616\,1& \ \ $-$10.972\,600\,9& \ \ $-$10.972\,590\,4& \ \ $-$0.456& \ \ $-$0.771 \\
\newpage
105(A)& \ \ 12\,952.921\,317\,7& \ \ 0.211& \ \ $-$9.955\,291\,3& \ \ $-$9.955\,266\,1& \ \ $-$9.955\,255\,4& \ \ $-$0.756& \ \ $-$1.078 \\
105(B)& \ \ 12\,952.924\,384\,8& \ \ 0.293& \ \ $-$9.952\,224\,2& \ \ $-$9.952\,202\,8& \ \ $-$9.952\,192\,1& \ \ $-$0.642& \ \ $-$0.964 \\
110(A)& \ \ 12\,953.805\,640\,4& \ \ 0.312& \ \ $-$9.070\,968\,6& \ \ $-$9.070\,889\,5& \ \ $-$9.070\,878\,6& \ \ $-$2.372& \ \ $-$2.699 \\
110(B)& \ \ 12\,953.808\,753\,7& \ \ 0.290& \ \ $-$9.067\,855\,3& \ \ $-$9.067\,779\,5& \ \ $-$9.067\,768\,6& \ \ $-$2.273& \ \ $-$2.600 \\
115(A)& \ \ 12\,954.577\,244\,8& \ \ 0.253& \ \ $-$8.299\,364\,2& \ \ $-$8.299\,354\,0& \ \ $-$8.299\,342\,9& \ \ $-$0.306& \ \ $-$0.638 \\
115(B)& \ \ 12\,954.580\,355\,1& \ \ 0.401& \ \ $-$8.296\,253\,9& \ \ $-$8.296\,202\,6& \ \ $-$8.296\,191\,5& \ \ $-$1.538& \ \ $-$1.870 \\

\bottomrule
\label{tab:meas_fr_Np2}

\end{longtable}

\begin{longtable}{lrrrrrrr}

\caption{Experimental positions (in wavenumber units cm$^{-1}$) of the measured fine-structure components of the $n$f4$(J=1)$ Rydberg states of H$_2$ relative to the GK(0,2) state, their statistical uncertainties (in MHz), their binding energies determined using the measured positions and the fitted $\gamma_{\textrm{SR}}^{N^+=4}$ and X$^+ (0,4) - $GK$(0,2)$ interval, their binding energies calculated by MQDT, their binding energies calculated by MQDT after correcting with a two-channel interaction model, the deviation between experiment and MQDT calculations, and the deviation between experiment and MQDT after correcting with a two-channel interaction model. The states are labelled by the principal quantum number $n$ and by the letter A, B or C to denote the different fine-structure components of $J$=1 in the order of increasing energy at each value of $n$.} \\

\toprule
$n$(Label) & \ \  $\tilde\nu_{\textrm{obs}}$ (cm$^{-1}$) & \ \  \ \ $\sigma$ (MHz) & \ \  $E_{\textrm{bind}} ^{\textrm{obs}} \ / \ hc$ (cm$^{-1}$) & \ \  $E_{\textrm{bind}} ^{\textrm{MQDT}} \ / \ hc$ (cm$^{-1}$) 
& \ \  $E_{\textrm{bind}} ^{\textrm{corr}} \ / \ hc$ (cm$^{-1}$) 
& \ \  $\nu_{\textrm{obs}} - \nu_{\textrm{MQDT}}$ (MHz) 
& \ \   $\nu_{\textrm{obs}} - \nu_{\textrm{corr}}$ (MHz)  \\
\midrule

33(A)&13\,198.368\,193\,6&0.737&$-$100.844\,879\,6&$-$100.845\,411\,8&$-$100.844\,854\,4&15.954&$-$0.757 \\
33(B)&13\,198.373\,337\,9&0.445&$-$100.839\,735\,3&$-$100.840\,320\,2&$-$100.839\,762\,8&17.534&0.824 \\
34(A)&13\,204.211\,198\,0&0.461&$-$95.001\,875\,2&$-$95.002\,825\,1&$-$95.001\,888\,1&28.476&0.386 \\
34(B)&13\,204.216\,385\,8&0.342&$-$94.996\,687\,4&$-$94.997\,718\,3&$-$94.996\,781\,4&30.904&2.816 \\
36(A)&13\,214.493\,183\,9&1.062&$-$84.719\,889\,3&$-$84.718\,918\,9&$-$84.719\,906\,9&$-$29.093&0.527 \\
36(B)&13\,214.498\,203\,3&0.489&$-$84.714\,869\,9&$-$84.713\,895\,1&$-$84.714\,883\,0&$-$29.225&0.393 \\
\newpage
39(A)&13\,227.025\,080\,8&0.261&$-$72.187\,992\,4&$-$72.187\,779\,5&$-$72.187\,935\,5&$-$6.385&$-$1.707 \\
39(B)&13\,227.029\,902\,3&0.212&$-$72.183\,170\,9&$-$72.182\,990\,4&$-$72.183\,146\,4&$-$5.413&$-$0.736 \\
42(A)&13\,236.971\,877\,7&0.296&$-$62.241\,195\,5&$-$62.241\,087\,7&$-$62.241\,147\,7&$-$3.234&$-$1.434 \\
42(B)&13\,236.976\,542\,1&0.282&$-$62.236\,531\,1&$-$62.236\,452\,2&$-$62.236\,512\,2&$-$2.368&$-$0.568 \\
44(A)&13\,242.503\,038\,3&0.574&$-$56.710\,035\,0&$-$56.709\,942\,6&$-$56.709\,977\,4&$-$2.769&$-$1.727 \\
44(B)&13\,242.507\,589\,6&0.345&$-$56.705\,483\,7&$-$56.705\,428\,7&$-$56.705\,463\,4&$-$1.648&$-$0.606 \\
45(A)&13\,244.996\,095\,8&0.414&$-$54.216\,977\,5&$-$54.216\,846\,2&$-$54.216\,872\,7&$-$3.936&$-$3.140 \\
45(B)&13\,245.000\,618\,7&0.254&$-$54.212\,454\,6&$-$54.212\,399\,6&$-$54.212\,426\,1&$-$1.648&$-$0.853 \\
46(A)&13\,247.328\,547\,4&0.414&$-$51.884\,525\,9&$-$51.884\,610\,9&$-$51.884\,631\,0&2.549&3.151 \\
46(B)&13\,247.332\,846\,0&0.532&$-$51.880\,227\,3&$-$51.880\,236\,3&$-$51.880\,256\,4&0.270&0.873 \\
47(A)&13\,249.513\,376\,2&0.270&$-$49.699\,697\,1&$-$49.699\,692\,2&$-$49.699\,707\,1&$-$0.146&0.302 \\
47(B)&13\,249.517\,664\,4&0.244&$-$49.695\,408\,9&$-$49.695\,394\,0&$-$49.695\,408\,9&$-$0.446&0.002 \\
48(A)&13\,251.563\,094\,9&0.692&$-$47.649\,978\,4&$-$47.649\,940\,4&$-$47.649\,951\,2&$-$1.139&$-$0.814 \\
48(B)&13\,251.567\,381\,0&0.530&$-$47.645\,692\,3&$-$47.645\,723\,3&$-$47.645\,734\,1&0.930&1.254 \\
48(C)&13\,251.573\,778\,3&0.840&$-$47.639\,295\,0&$-$47.639\,321\,9&$-$47.639\,332\,7&0.807&1.131 \\
50(A)&13\,255.299\,723\,8&0.292&$-$43.913\,349\,5&$-$43.913\,328\,1&$-$43.913\,332\,7&$-$0.641&$-$0.502 \\
50(B)&13\,255.303\,770\,1&0.284&$-$43.909\,303\,2&$-$43.909\,286\,8&$-$43.909\,291\,4&$-$0.491&$-$0.352 \\
50(C)&13\,255.309\,559\,8&0.717&$-$43.903\,513\,5&$-$43.903\,517\,6&$-$43.903\,522\,2&0.123&0.262 \\
51(A)&13\,257.005\,306\,2&0.568&$-$42.207\,767\,1&$-$42.207\,740\,0&$-$42.207\,742\,3&$-$0.812&$-$0.743 \\
51(B)&13\,257.009\,281\,9&0.453&$-$42.203\,791\,4&$-$42.203\,792\,8&$-$42.203\,795\,1&0.042&0.112 \\
51(C)&13\,257.014\,810\,9&0.770&$-$42.198\,262\,4&$-$42.198\,272\,4&$-$42.198\,274\,7&0.300&0.369 \\
52(A)&13\,258.613\,443\,8&0.353&$-$40.599\,629\,5&$-$40.599\,628\,6&$-$40.599\,628\,9&$-$0.027&$-$0.016 \\
\newpage
52(B)&13\,258.617\,297\,6&0.253&$-$40.595\,775\,7&$-$40.595\,779\,1&$-$40.595\,779\,4&0.102&0.113 \\
52(C)&13\,258.622\,621\,5&0.360&$-$40.590\,451\,8&$-$40.590\,468\,1&$-$40.590\,468\,4&0.489&0.499 \\
55(A)&13\,262.922\,521\,7&0.365&$-$36.290\,551\,6&$-$36.290\,539\,9&$-$36.290\,535\,9&$-$0.351&$-$0.469 \\
55(B)&13\,262.926\,066\,2&0.349&$-$36.287\,007\,1&$-$36.286\,999\,6&$-$36.286\,995\,6&$-$0.225&$-$0.343 \\
55(C)&13\,262.930\,965\,1&0.643&$-$36.282\,108\,2&$-$36.282\,117\,9&$-$36.282\,113\,9&0.291&0.172 \\
58(A)&13\,266.580\,261\,6&0.333&$-$32.632\,811\,7&$-$32.632\,810\,4&$-$32.632\,803\,7&$-$0.039&$-$0.240 \\
58(B)&13\,266.583\,487\,1&0.348&$-$32.629\,586\,2&$-$32.629\,589\,9&$-$32.629\,583\,2&0.111&$-$0.090 \\
58(C)&13\,266.588\,179\,7&0.534&$-$32.624\,893\,6&$-$32.624\,913\,1&$-$32.624\,906\,4&0.585&0.384 \\
60(A)&13\,268.719\,868\,6&0.423&$-$30.493\,204\,7&$-$30.493\,213\,6&$-$30.493\,205\,6&0.267&0.027 \\
60(B)&13\,268.722\,874\,5&0.494&$-$30.490\,198\,8&$-$30.490\,204\,0&$-$30.490\,196\,0&0.156&$-$0.084 \\
60(C)&13\,268.727\,493\,7&0.532&$-$30.485\,579\,6&$-$30.485\,579\,3&$-$30.485\,571\,3&$-$0.009&$-$0.249 \\
64(A)&13\,272.412\,890\,1&0.740&$-$26.800\,183\,2&$-$26.800\,192\,3&$-$26.800\,182\,5&0.273&$-$0.021 \\
64(B)&13\,272.415\,493\,5&0.458&$-$26.797\,579\,8&$-$26.797\,581\,6&$-$26.797\,571\,8&0.054&$-$0.240 \\
64(C)&13\,272.420\,146\,5&0.662&$-$26.792\,926\,8&$-$26.792\,939\,5&$-$26.792\,929\,7&0.381&0.086 \\
67(A)&13\,274.759\,445\,6&0.736&$-$24.453\,627\,7&$-$24.453\,650\,6&$-$24.453\,639\,9&0.686&0.365 \\
67(B)&13\,274.761\,771\,9&1.193&$-$24.451\,301\,4&$-$24.451\,310\,8&$-$24.451\,300\,1&0.282&$-$0.040 \\
67(C)&13\,274.766\,496\,1&0.457&$-$24.446\,577\,2&$-$24.446\,596\,7&$-$24.446\,586\,0&0.584&0.263 \\
70(A)&13\,276.810\,737\,5&0.415&$-$22.402\,335\,8&$-$22.402\,355\,9&$-$22.402\,344\,5&0.602&0.262 \\
70(B)&13\,276.812\,853\,3&1.345&$-$22.400\,220\,0&$-$22.400\,259\,6&$-$22.400\,248\,2&1.187&0.846 \\
70(C)&13\,276.817\,641\,1&0.299&$-$22.395\,432\,2&$-$22.395\,451\,4&$-$22.395\,440\,0&0.575&0.235 \\
80(A)&13\,282.061\,468\,2&0.132&$-$17.151\,605\,1&$-$17.151\,583\,1&$-$17.151\,570\,5&$-$0.660&$-$1.036 \\
80(B)&13\,282.062\,936\,9&0.144&$-$17.150\,136\,4&$-$17.150\,115\,7&$-$17.150\,103\,1&$-$0.621&$-$0.997 \\
\newpage
80(C)&13\,282.068\,060\,3&0.123&$-$17.145\,013\,0&$-$17.144\,974\,8&$-$17.144\,962\,2&$-$1.146&$-$1.522 \\
90(A)&13\,285.661\,120\,3&0.534&$-$13.551\,953\,0&$-$13.551\,936\,1&$-$13.551\,923\,0&$-$0.507&$-$0.899 \\
90(B)&13\,285.662\,171\,2&0.468&$-$13.550\,902\,1&$-$13.550\,882\,1&$-$13.550\,869\,0&$-$0.600&$-$0.991 \\
90(C)&13\,285.667\,541\,5&0.225&$-$13.545\,531\,8&$-$13.545\,474\,1&$-$13.545\,461\,0&$-$1.730&$-$2.122 \\
100(A)&13\,288.235\,774\,2&0.176&$-$10.977\,299\,1&$-$10.977\,279\,8&$-$10.977\,266\,5&$-$0.579&$-$0.978 \\
100(B)&13\,288.236\,551\,5&0.367&$-$10.976\,521\,8&$-$10.976\,501\,4&$-$10.976\,488\,1&$-$0.612&$-$1.011 \\
100(C)&13\,288.242\,153\,6&0.152&$-$10.970\,919\,7&$-$10.970\,900\,2&$-$10.970\,886\,9&$-$0.585&$-$0.984 \\
110(A)&13\,290.140\,625\,2&0.340&$-$9.072\,448\,1&$-$9.072\,415\,0&$-$9.072\,401\,6&$-$0.993&$-$1.395 \\
110(B)&13\,290.141\,208\,0&0.576&$-$9.071\,865\,3&$-$9.071\,825\,5&$-$9.071\,812\,1&$-$1.194&$-$1.596 \\
110(C)&13\,290.146\,962\,5&0.230&$-$9.066\,110\,8&$-$9.066\,086\,4&$-$9.066\,073\,0&$-$0.732&$-$1.134 \\

\bottomrule
\label{tab:meas_fr_Np4}

\end{longtable}

\begin{longtable}{lrrrrrrr}

\caption{Experimental positions (in wavenumber units cm$^{-1}$) of the measured fine-structure components of the $n$f6$(J=3)$ Rydberg states of H$_2$ relative to the I$^+$(0,4) state, their statistical uncertainties (in MHz), their binding energies determined using the measured positions and the fitted $\gamma_{\textrm{SR}}^{N^+=6}$ and X$^+ (0,6) - $ I$^+$(0,4) interval, their binding energies calculated by MQDT, their binding energies calculated by MQDT after correcting with a two-channel interaction model, the deviation between experiment and MQDT calculations, and the deviation between experiment and MQDT after correcting with a two-channel interaction model. The states are labelled by the principal quantum number $n$ and by the letter A, B or C to denote the different fine-structure components of $J$=3 in the order of increasing energy at each value of $n$.} \\

\toprule
$n$(Label) & \ \  $\tilde\nu_{\textrm{obs}}$ (cm$^{-1}$) & \ \  \ \ $\sigma$ (MHz) & \ \  $E_{\textrm{bind}} ^{\textrm{obs}} \ / \ hc$ (cm$^{-1}$) & \ \  $E_{\textrm{bind}} ^{\textrm{MQDT}} \ / \ hc$ (cm$^{-1}$) 
& \ \  $E_{\textrm{bind}} ^{\textrm{corr}} \ / \ hc$ (cm$^{-1}$) 
& \ \  $\nu_{\textrm{obs}} - \nu_{\textrm{MQDT}}$ (MHz) 
& \ \   $\nu_{\textrm{obs}} - \nu_{\textrm{corr}}$ (MHz)  \\
\midrule

28(A)&12\,765.093\,829\,9&0.736&$-$140.090\,312\,0&$-$140.090\,689\,2&$-$140.090\,074\,8&11.307&$-$7.111 \\
28(B)&12\,765.102\,112\,1&0.613&$-$140.082\,029\,8&$-$140.082\,486\,2&$-$140.081\,871\,9&13.681&$-$4.735 \\
\newpage
29(A)&12\,774.589\,448\,0&0.770&$-$130.594\,693\,9&$-$130.595\,462\,6&$-$130.594\,419\,8&23.044&$-$8.219 \\
29(B)&12\,774.597\,695\,4&0.638&$-$130.586\,446\,5&$-$130.587\,275\,6&$-$130.586\,232\,9&24.854&$-$6.405 \\
31(A)&12\,790.917\,370\,9&0.843&$-$114.266\,771\,1&$-$114.265\,832\,3&$-$114.266\,689\,8&$-$28.143&$-$2.436 \\
31(B)&12\,790.925\,553\,9&0.703&$-$114.258\,588\,1&$-$114.257\,708\,0&$-$114.258\,565\,4&$-$26.383&$-$0.679 \\
33(A)&12\,804.349\,271\,8&0.754&$-$100.834\,870\,2&$-$100.834\,558\,4&$-$100.834\,767\,0&$-$9.347&$-$3.094 \\
33(B)&12\,804.357\,388\,3&0.588&$-$100.826\,753\,7&$-$100.826\,478\,9&$-$100.826\,687\,4&$-$8.237&$-$1.986 \\
36(A)&12\,820.459\,641\,9&0.882&$-$84.724\,500\,1&$-$84.724\,342\,1&$-$84.724\,408\,6&$-$4.736&$-$2.742 \\
36(B)&12\,820.467\,665\,8&0.624&$-$84.716\,476\,2&$-$84.716\,336\,6&$-$84.716\,403\,1&$-$4.185&$-$2.190 \\
40(A)&12\,836.561\,866\,2&0.999&$-$68.622\,275\,8&$-$68.622\,183\,6&$-$68.622\,204\,1&$-$2.764&$-$2.150 \\
40(B)&12\,836.569\,742\,8&0.890&$-$68.614\,399\,2&$-$68.614\,314\,3&$-$68.614\,334\,8&$-$2.545&$-$1.932 \\
43(A)&12\,845.805\,454\,4&1.292&$-$59.378\,687\,6&$-$59.378\,608\,5&$-$59.378\,616\,7&$-$2.371&$-$2.125 \\
43(B)&12\,845.813\,251\,8&0.983&$-$59.370\,890\,2&$-$59.370\,890\,4&$-$59.370\,898\,6&0.006&0.252 \\
46(A)&12\,853.299\,748\,0&1.388&$-$51.884\,393\,9&$-$51.884\,383\,5&$-$51.884\,385\,7&$-$0.312&$-$0.247 \\
46(B)&12\,853.307\,181\,6&0.986&$-$51.876\,960\,3&$-$51.876\,889\,1&$-$51.876\,891\,3&$-$2.135&$-$2.070 \\
49(A)&12\,859.459\,748\,2&1.074&$-$45.724\,393\,7&$-$45.724\,412\,2&$-$45.724\,411\,2&0.554&0.523 \\
49(B)&12\,859.466\,877\,4&1.079&$-$45.717\,264\,5&$-$45.717\,254\,2&$-$45.717\,253\,2&$-$0.309&$-$0.341 \\
52(A)&12\,864.584\,319\,7&1.057&$-$40.599\,822\,2&$-$40.599\,789\,7&$-$40.599\,786\,9&$-$0.975&$-$1.060 \\
52(B)&12\,864.591\,028\,2&1.331&$-$40.593\,113\,7&$-$40.593\,112\,3&$-$40.593\,109\,5&$-$0.043&$-$0.127 \\
52(C)&12\,864.595\,232\,7&1.619&$-$40.588\,909\,2&$-$40.588\,894\,0&$-$40.588\,891\,2&$-$0.456&$-$0.541 \\
55(A)&12\,868.893\,320\,7&0.666&$-$36.290\,821\,2&$-$36.290\,854\,2&$-$36.290\,850\,4&0.989&0.873 \\
55(B)&12\,868.899\,286\,6&0.990&$-$36.284\,855\,3&$-$36.284\,780\,3&$-$36.284\,776\,5&$-$2.249&$-$2.364 \\
55(C)&12\,868.903\,441\,0&0.945&$-$36.280\,700\,9&$-$36.280\,728\,6&$-$36.280\,724\,8&0.830&0.714 \\
\newpage
58(A)&12\,872.550\,985\,2&1.067&$-$32.633\,156\,7&$-$32.633\,252\,3&$-$32.633\,247\,9&2.865&2.732 \\
58(B)&12\,872.556\,432\,0&1.956&$-$32.627\,709\,9&$-$32.627\,828\,2&$-$32.627\,823\,8&3.546&3.413 \\
58(C)&12\,872.560\,573\,6&1.238&$-$32.623\,568\,3&$-$32.623\,581\,9&$-$32.623\,577\,5&0.407&0.274 \\
61(A)&12\,875.682\,197\,4&0.798&$-$29.501\,944\,5&$-$29.502\,032\,2&$-$29.502\,027\,4&2.628&2.486 \\
61(B)&12\,875.686\,921\,2&2.190&$-$29.497\,220\,7&$-$29.497\,233\,1&$-$29.497\,228\,3&0.371&0.228 \\
61(C)&12\,875.691\,579\,8&0.822&$-$29.492\,562\,1&$-$29.492\,635\,7&$-$29.492\,630\,9&2.206&2.063 \\
65(A)&12\,879.201\,621\,1&0.969&$-$25.982\,520\,8&$-$25.982\,479\,5&$-$25.982\,474\,5&$-$1.239&$-$1.388 \\
65(B)&12\,879.205\,652\,0&2.746&$-$25.978\,489\,9&$-$25.978\,418\,5&$-$25.978\,413\,5&$-$2.141&$-$2.290 \\
65(C)&12\,879.210\,744\,1&1.423&$-$25.973\,397\,8&$-$25.973\,300\,3&$-$25.973\,295\,3&$-$2.924&$-$3.073 \\
70(A)&12\,882.781\,119\,1&1.199&$-$22.403\,022\,8&$-$22.403\,130\,0&$-$22.403\,125\,0&3.213&3.062 \\
70(B)&12\,882.784\,417\,9&2.572&$-$22.399\,724\,0&$-$22.399\,822\,5&$-$22.399\,817\,5&2.952&2.801 \\
70(C)&12\,882.789\,977\,4&1.296&$-$22.394\,164\,5&$-$22.394\,105\,2&$-$22.394\,100\,2&$-$1.779&$-$1.930 \\
80(A)&12\,888.031\,603\,3&0.417&$-$17.152\,538\,6&$-$17.152\,506\,7&$-$17.152\,501\,8&$-$0.958&$-$1.106 \\
80(B)&12\,888.033\,817\,0&0.357&$-$17.150\,324\,9&$-$17.150\,252\,9&$-$17.150\,248\,0&$-$2.160&$-$2.308 \\
80(C)&12\,888.040\,484\,1&0.423&$-$17.143\,657\,8&$-$17.143\,632\,7&$-$17.143\,627\,8&$-$0.754&$-$0.902 \\
90(A)&12\,891.631\,177\,4&0.671&$-$13.552\,964\,5&$-$13.552\,951\,1&$-$13.552\,946\,3&$-$0.403&$-$0.547 \\
90(C)&12\,891.639\,953\,9&0.618&$-$13.544\,188\,0&$-$13.544\,147\,1&$-$13.544\,142\,3&$-$1.227&$-$1.371 \\
95(A)&12\,893.020\,024\,8&0.878&$-$12.164\,117\,1&$-$12.164\,101\,6&$-$12.164\,096\,9&$-$0.466&$-$0.608 \\
95(C)&12\,893.028\,793\,1&0.606&$-$12.155\,348\,8&$-$12.155\,319\,3&$-$12.155\,314\,6&$-$0.886&$-$1.027 \\
100(A)&12\,894.205\,758\,5&0.690&$-$10.978\,383\,4&$-$10.978\,353\,7&$-$10.978\,349\,0&$-$0.892&$-$1.031 \\
100(C)&12\,894.214\,535\,5&0.399&$-$10.969\,606\,4&$-$10.969\,588\,1&$-$10.969\,583\,4&$-$0.550&$-$0.690 \\

\bottomrule
\label{tab:meas_fr_Np6}

\end{longtable}